\documentclass[11pt]{article}

\usepackage{amsfonts}
\usepackage{epsfig}
\usepackage{latexsym}
\usepackage{amsmath}
\usepackage{amssymb}
\usepackage{float}

\def\bequ{\begin{equation}}
\def\eequ{\end{equation}}
\def\barr{\begin{array}}
\def\earr{\end{array}}
\def\half{{1\over 2}}
\def\ben{\begin{equation}}
\def\een{\end{equation}}
\def\bena{\begin{eqnarray}}
\def\eena{\end{eqnarray}}

\def\b1{e^0}

\newcommand{\be}{\begin{equation}}
\newcommand{\ee}{\end{equation}}
\def\bea{\begin{eqnarray}}
\def\eea{\end{eqnarray}}

\def\half {{1 \over 2}}

\def\be{\begin{equation}}
\def\ee{\end{equation}}
\def\bea{\begin{eqnarray}}
\def\eea{\end{eqnarray}}

\def\lesssim{\mathrel{\hbox{\rlap{\hbox{\lower4pt\hbox{$\sim$}}}\hbox{$<$}}}}
\def\gtrsim{\mathrel{\hbox{\rlap{\hbox{\lower4pt\hbox{$\sim$}}}\hbox{$>$}}}}

\begin{document}
\title{{\LARGE \bf{Global embedding of the Kerr black hole \\ event horizon into hyperbolic 3-space}}}
\author{
G. W. Gibbons$^{1,}$\footnote{G.W.Gibbons@damtp.cam.ac.uk}~, \
C. A. R. Herdeiro$^{2,}$\footnote{crherdei@fc.up.pt}~ \ and
 C. Rebelo$^{2,}$\footnote{mrebelo@fc.up.pt}
\\
\\ {$^{1}${\em D.A.M.T.P.}}
\\ {\em  Cambridge University}
\\ {\em Wilberforce Road, Cambridge CB3 0WA, U.K.}
\\
\\ $^{2}${\em Departamento de F\'\i sica e Centro de F\'\i sica do Porto}
\\ {\em Faculdade de Ci\^encias da Universidade do Porto}
\\ {\em Rua do Campo Alegre, 687,  4169-007 Porto, Portugal}}

\date{\today}       
 \maketitle

\begin{abstract}
An explicit global and unique isometric embedding into hyperbolic 3-space, $H^3$, of an axi-symmetric 2-surface with Gaussian curvature bounded below is given. In particular, this allows the embedding into $H^3$ of surfaces of revolution having negative, but finite, Gaussian curvature at smooth fixed points of the $U(1)$ isometry. As an example, we exhibit the global embedding of the Kerr-Newman event horizon into $H^3$, for arbitrary values of the angular momentum. For this example, considering a quotient of $H^3$ by the Picard group, we show that the hyperbolic embedding fits in a fundamental domain of the group up to a slightly larger value of the angular momentum than the limit for which a global embedding into Euclidean 3-space is possible. An embedding of the double-Kerr event horizon is also presented, as an example of an embedding which cannot be made global. 
\end{abstract}

\section{Introduction}
Since Gauss proved his famous  {\it Theorema Egregium}, it has become customary to make a clear distinction  between the intrinsic properties of a closed surface $S$,
whether local quantities such as  its  Gauss curvature 
\bequ
K= \frac{1}{R_1 R_2} \ , \ \ \ \ \ {\rm where \ R_1,R_2 \  
are \ the \ principal \ radii \ of \ curvature} , \eequ
or non-local quantities such as its  area $A$, 
which depend only on the  metric, and  extrinsic properties such as its mean curvature 
\bequ H=\half \left( {1 \over R_1} + {1 \over R_2} \right) \ , \eequ which is local, or its largest diameter $D$, which is non-local, since these  may depend on which particular embedding one adopts.

However, this distinction may break down
if the embedding in question is {\it rigid}, that is unique.
By  theorems of Weyl, Pogorelov and others 
(see \cite{Hopf, Nirenberg}), this is 
true of smooth 2-surfaces whose Gauss  curvature
is everywhere positive; then there exists, up to isometries,
a unique isometric embedding into Euclidean 3-space ${\Bbb E} ^3$. 
Since this embedding 
${\bf x } = {\bf x}( u^i )$  is given by three   highly non-linear      
partial differential equations depending on the intrinsic metric,
\ben
g_{ij} = \partial _i{\bf x}  .\partial _j {\bf x}  \ , 
\een the extrinsic quantities depend in a  highly non-local
fashion on  the metric.
 
If the Gauss curvature is somewhere negative
then $S$ may  or may not admit a global isometric embedding into
${\Bbb E} ^3$. In \cite{Frolov} an example of a 2-surface with a patch of negative curvature that can be globally embedded into ${\Bbb E} ^3$ is given (another simple example is a ``donut" like surface); in the same reference it is also shown that if the 2-surface admits a $U(1)$ action and the Gaussian curvature is negative at a smooth fixed point of such action, then a vicinity of the fixed point cannot be embedded, not even locally, into ${\Bbb E} ^3$. If the fixed point having negative Gaussian curvature is not smooth, however, a global embedding may exist. An example is the Kerr ergo-sphere \cite{Pelavas:2000za}.

 But if the Gauss curvature is bounded below
by some negative constant,  
\ben
K \ge - { 1 \over L ^2}  \,,
\een 
then Pogorelov \cite{Pogorelov} has shown that there is always an
isometric embedding, unique up to isometries,  
into hyperbolic space $H^3$  with radius of curvature $L$. 
By scaling $L$ may always be taken to be unity.

These facts have clear relevance to general relativity, both at the 
 classical and
the quantum level:  
\begin{itemize}
\item Historically (e.g. \cite{Flamm,Smarr,Friedman,Rosen}; see also \cite{Goenner}) isometric embeddings have frequently
been used to gain intuition about geometric features of spacetimes.  
If the embedding is rigid, then such intuition is less likely
to be misleading than if the embedding is flexible, and
many equivalent embeddings exist. For an interesting
account  of flexible non-convex polyhedra see \cite{Stewart}.   

\item In  recent efforts to define a so-called  
quasi-local mass \cite{Yau},
functional associated with a two surface $S$,  embedded in a 
four-dimensional Lorentzian spacetime, use has been made of 
the Pogorolev theorem to embed $S$ isometrically into $H^3$ and hence
considering $H^3$ as points equidistant to the future from some origin (i.e the mass shell),
in flat Minkowski spacetime ${\Bbb E} ^{3,1}$.      
The embedding into ${\Bbb E} ^{3,1}$ is not expected to be unique.

\item In a recent paper \cite{Gibbons}, one of us gave an intrinsic
formulation of Thorne's Hoop conjecture, and used  as a technical tool
an isometric embedding into ${\Bbb E} ^3$ to prove it
in some   particular cases.
 \end{itemize}

In this paper, we shall, with these motivations in mind,
provide explicit  isometric  embeddings
of the horizons of Kerr-Newman and double Kerr-Newman
black holes into $H^3$. The former is global. The latter is
local,  because of the presence of a strut which is 
required to keep the two black holes
apart.  

An isometric embedding of the Kerr-Newman horizon into
${\Bbb E} ^3$ was first given by Smarr \cite{Smarr} 
who discovered that for $J > \sqrt{3}M^2/2$ the Gauss curvature $K$ at the
north and south poles becomes negative and a global  isometric embedding   
into ${\Bbb E} ^3$ is no longer possible. A local embedding into
3-dimensional Minkowski spacetime ${\Bbb E} ^{2,1}$  is possible 
but this is  not global \cite{GHWW}.  To circumvent
this problem Frolov \cite{Frolov} has given a global embedding 
into four-dimensional  Euclidean space ${\Bbb E} ^4$. 
It is not known whether the Frolov embedding is rigid, but it
seems unlikely.

\section{The upper half space model}   
\label{uhsmodel}

Events $(T,X,Y,Z)$ in ${\Bbb E}^{3,1}$ are in one-to-one correspondence
with Hermitian two by two matrices 
\ben
{\bf X} = \left(\begin{matrix}T+Z & X+iY \cr X-iY& T-Z \end{matrix}\right) \,, 
\een
and the Lorentz group $SO(3,1)= PSL(2,{\Bbb C})$ acts as
\ben
{\bf X} \rightarrow S {\bf X} S^\dagger\,,\qquad {\bf X} \in SL(2,{\Bbb C}) \,.
\een
Translations act as matrix addition:
\ben
X \rightarrow X + A\,. \label{translations}
\een
Hyperbolic space $H^3$  corresponds to events  
\ben
{\bf X} = \left(\begin{matrix} \displaystyle{1 \over z} &  \displaystyle{x+iy \over z} \cr \displaystyle{x-iy \over z} & 
\displaystyle{ {x^2 + y^2 \over z}+z} \end{matrix}\right) \,, \qquad z>0\,, \  x+iy \in {\Bbb C} \,. \een
To embed $H^3$ into Minkowski spacetime ${\Bbb E}^{3,1}$ 
one simply equates the two matrix formulas above. The induced metric on the surface 
\bequ 
X^2+Y^2+Z^2-T^2=-1\ , \label{hypersurface} \eequ 
is then
\ben
ds ^2 = {1 \over z^2} \bigl(d z^2 + d x^2 + dy ^2 \bigr ) \,. \label{uhsmetric}
\een
It is now simple to embed any surface in the upper half space model of $H^3$ \eqref{uhsmetric} into  
${\Bbb E}^{3,1}$. 

\section{Identifications}
\label{identifications}

Rather than embedding into $H^3$, one might 
consider  making  identifications under the action of some discrete subgroup
 $\Gamma \subset SO(3,1)$.  
Quotients $H^3/\Gamma$  which  are closed 
or possibly merely of finite volume
are of interest in cosmology, in models where the spatial sections
of $k=-1$ Friedman-Lemaitre universes are taken to be of the
 so-called non-Euclidean honeycomb form $H^3/\Gamma$ \cite{CW}; they are also of interest in some Kaluza-Klein compactifications \cite{GZ,BGH-N}. The simplest non-compact  quotient of finite volume is obtained by taking  $\Gamma$ to be  the Picard group \cite{P}
$SL(2,{\Bbb G}) $, where ${\Bbb G}$ are the Gaussian integers
$m+in$, $m,n \in {\Bbb Z}$. This is a double quotient
$SL(2,{\Bbb G})\backslash SL(2,{\Bbb C})/SU(2)$.
This example has been used in cosmology for studies of the effects of global spatial  topology on the CMB \cite{C,D}.

A fundamental domain, completely analogous
to the fundamental domain of the modular group $SL(2,{\Bbb Z})$,  
acting on the upper half plane    is \cite{P} 
\ben
|x| \le \half\,,\qquad |y| \le \half \,,
\qquad x^2 + y^2 + z^2 \ge 1\,. \label{fundamentald}
\een 
Below we shall exhibit the embedding of the Kerr-Newman event horizon into $H^3$ and show that, up to a certain value of the angular momentum, it can be taken to lie completely  inside this fundamental domain.

There is an interesting connection with the  unique four-dimensional  
self-dual Lorentzian  lattice $\Pi^{3,1}$ \cite{GO}. The mathematics and physics literatures
differ somewhat in their use of the word lattice.
In the mathematical literature a {\it lattice} $\Lambda$  
is taken to be a finitely generated discrete abelian group.
Thinking of $\Lambda$  as a subgroup of ${\Bbb R}^n$ 
leads us to the toroidal quotient $ T^n= (S^1)^n ={\Bbb R}^n /\Lambda  $.
In the present case one may think of the lattice as 
matrices  $A$ in (\ref{translations}) whose entries
are Gaussian integers. 
In the physics literature a {\it lattice}  is usually taken to be 
a set of points in ${\Bbb R} ^n$ invariant under the action
of a lattice $\Lambda$. Typically these points are the orbits
in ${\Bbb R} ^n$ of a lattice $\Lambda$. 
A further source of confusion is that the underlying affine space 
${\Bbb R}^n$,
is frequently  endowed with a metric of signature $(s,t)$.
Thus, the mathematicians at least, should speak of a lattice
with metric, especially since the theory of the 
classification of such lattices
depends crucially on the signature of the metric.

In the present case $(s,t)=(3,1)$, corresponding  to restricting  
$(T,X,Y,Z)$ to take integer values \cite{GO}.
The point group of the lattice is $SO(3,1;{\Bbb Z})$ \cite{CW}  which
is covered by Picard's group $SL(2,{\Bbb G})$. 
The corresponding lattice, in the physical sense, was 
used by Schild in an attempt to model a discrete spacetime
\cite{S1,S2}. One can also think of the associated
quotient as a spacetime which is periodic both in space and in  time.
An interesting question is then whether one may embed our surfaces
into a unit cell of this lattice universe, which we shall address in the example of section \ref{kn}.

\section{General embedding formulae}
\label{generalformulae}
We shall now present the explicit embedding formulas for a 
2-surface admitting a $U(1)$ action in $H^3$. 
We start with the upper half space model for hyperbolic 3-space
$H^3$, discussed in section \ref{uhsmodel}:
\bequ
ds^2=\frac{L^2}{z^2}\left(dx^2+dy^2+dz^2\right)
=\frac{L^2}{z^2}\left(dr^2+r^2d\phi^2+dz^2\right)
\ , \eequ
where $L$ is the ``radius'' of the hyperbolic space, $z>0$ and in the
last equality we have used polar coordinates. We wish to construct a
global embedding of the
2-surface 
\bequ
ds^2=R^2\left(a^2(u)du^2+b^2(u)d\phi^2\right) \ , \eequ
in $H^3$. The functions $a(u),b(u)$ are dimensionless; $u$ is
a polar coordinate with range $-1\le u\le 1$, with equalities attained, respectively, at the ``south" and ``north" poles; $R$ is a length scale. The embedding functions
are 
\bequ
r(u)=\frac{b(u)z(u)}{k} \ , \qquad
\left(b(u)^2+k^2\right)\frac{z'(u)}{z(u)}=-b(u)b'(u)\pm
\sqrt{[a(u)b(u)]^2+k^2\left[a(u)^2-b'(u)^2\right]} \ ,
\eequ
where $k=L/R$. To make contact with the embedding used in \cite{Smarr} for the
Kerr-Newman horizon and in \cite{Costa:2009wj} for the double-Kerr
horizons we take $a(u)^2=1/b(u)^2=1/g(u)$; then, the embedding
functions are
\bequ
r(u)=\frac{\sqrt{g(u)}z(u)}{k} \ , \qquad
\left(g(u)+k^2\right)\frac{z'(u)}{z(u)}=-\frac{g'(u)}{2}\pm
\sqrt{1+\frac{k^2}{g(u)}\left[1-\frac{g'(u)^2}{4}\right]} \
. \label{embed} \eequ
In limit of large radius $k$ (i.e. flat space limit) these functions
reduce to 
\bequ
\tilde{r}(u)=R\sqrt{g(u)} \ , \qquad
\tilde{z}'(u)=\pm
\sqrt{\frac{R^2}{g(u)}\left[1-\frac{g'(u)^2}{4}\right]} \ , \eequ
where $\tilde{r}$ and $\tilde{z}$ are cylindrical polar coordinates in
$\mathbb{E}^3$. These are exactly the embedding functions used in
\cite{Costa:2009wj}. Thus, while the embedding in $\mathbb{E}^3$ will
fail when $g'(u)^2>4$, the same embedding in $H^3$ will be
possible if $g'(u)$ is bounded. The advantage of using hyperbolic space becomes
therefore manifest. Note that the Gaussian curvature of the 2-surface is
\bequ
K=-\frac{g''(u)}{2R^2} \ . \eequ
From this formula one can see that the failure of the embedding is not directly associated to the Gaussian curvature becoming negative. This is manifest in the Kerr-Newman example - see Fig. \ref{kerr}.

\section{The Kerr-Newman horizon}
\label{kn}
\begin{figure}[h!]
\centering\includegraphics[height=2in,width=3in]{{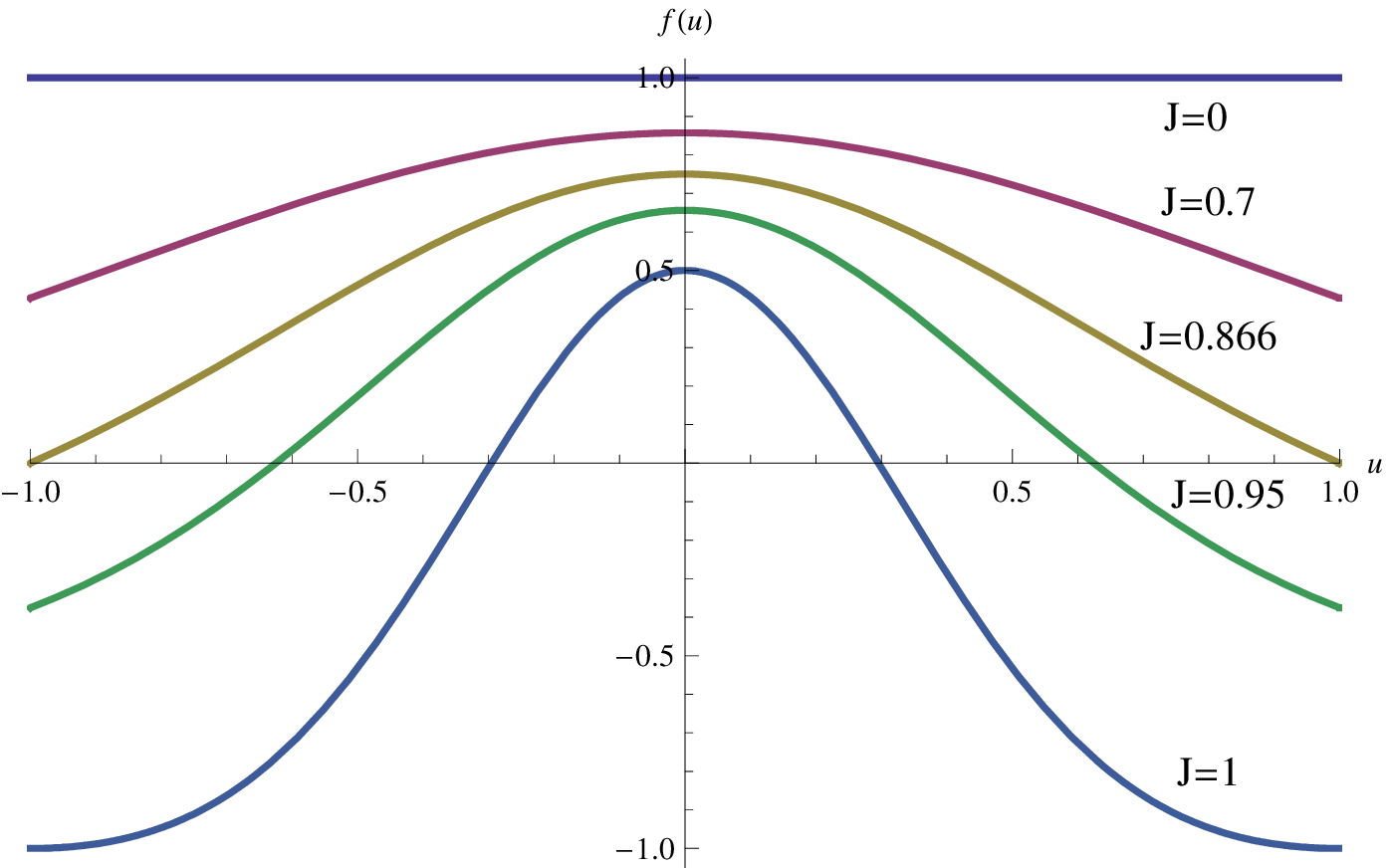}} 
\ \ \ \ \ \ 
\centering\includegraphics[height=2in,width=3in]{{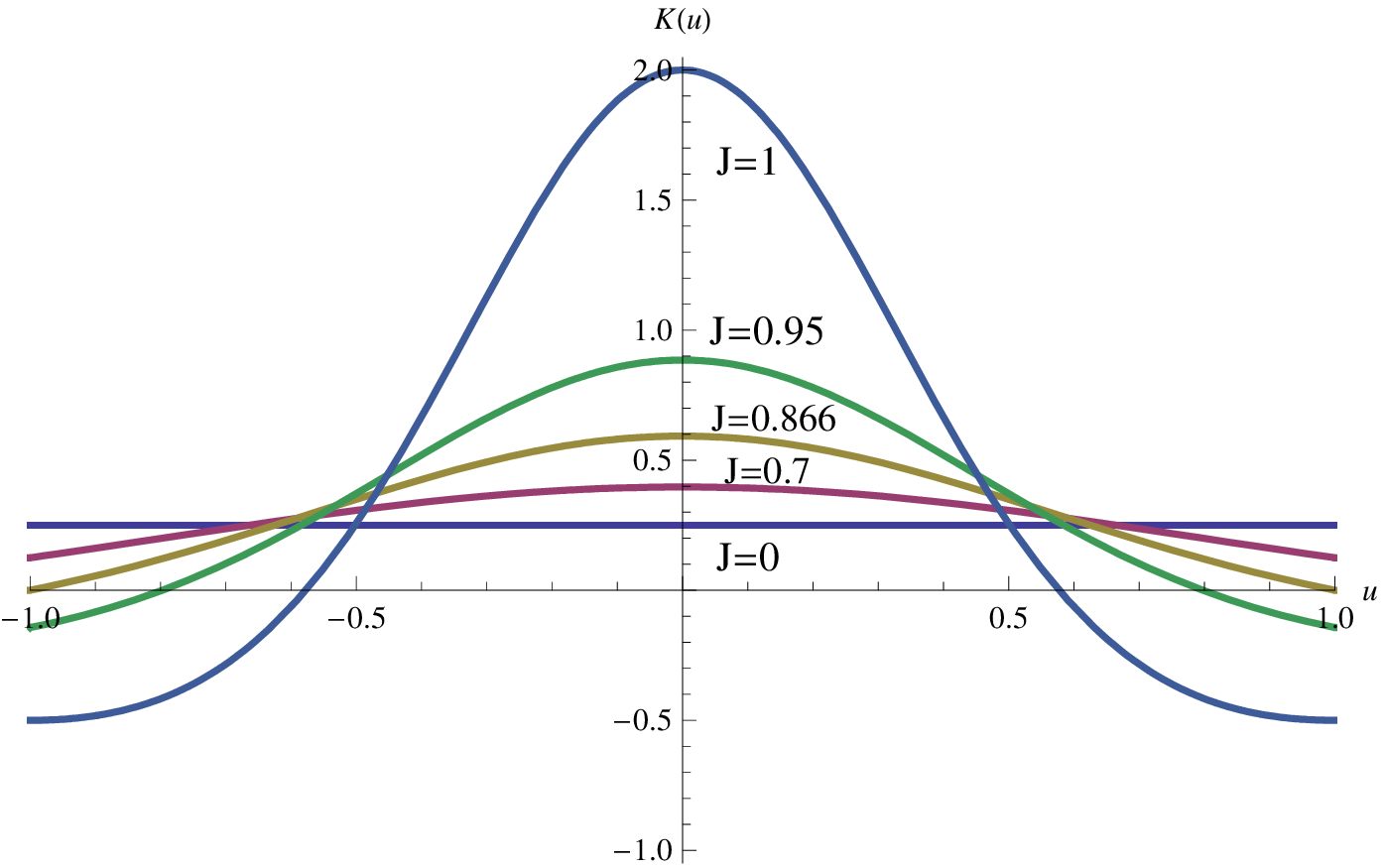}}
\begin{picture}(0,0)(0,0)
\end{picture}
\caption{$f(u)$ (left) and Gauss curvature (right) for the Kerr-Newman
 solution, for fixed mass $M=1$ and various values of the angular momentum $J$. The
 embedding in $\mathbb{E}^3$ fails when the $f(u)$ becomes
 negative; this function is bounded by $-1$ for any $J$. Note that there are regions where the curvature is positive but the embedding fails.}
\label{kerr}
\end{figure}

\begin{figure}[h!]
\centering\includegraphics[height=3in,width=1.5in]{{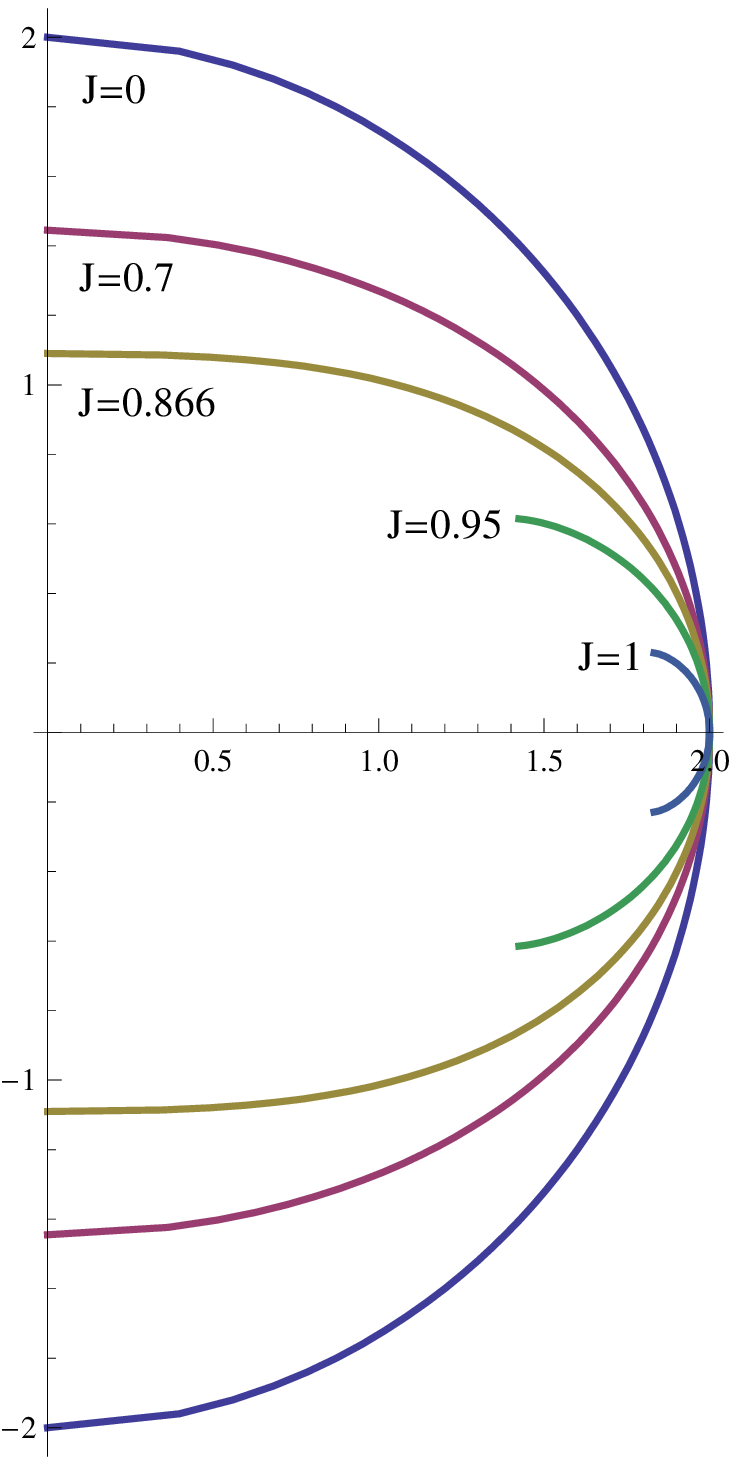}}
\ \ \ \ \ \ \ \ \ \ \ \  \ \ \ \ \ 
\centering\includegraphics[height=3in,width=1.5in]{{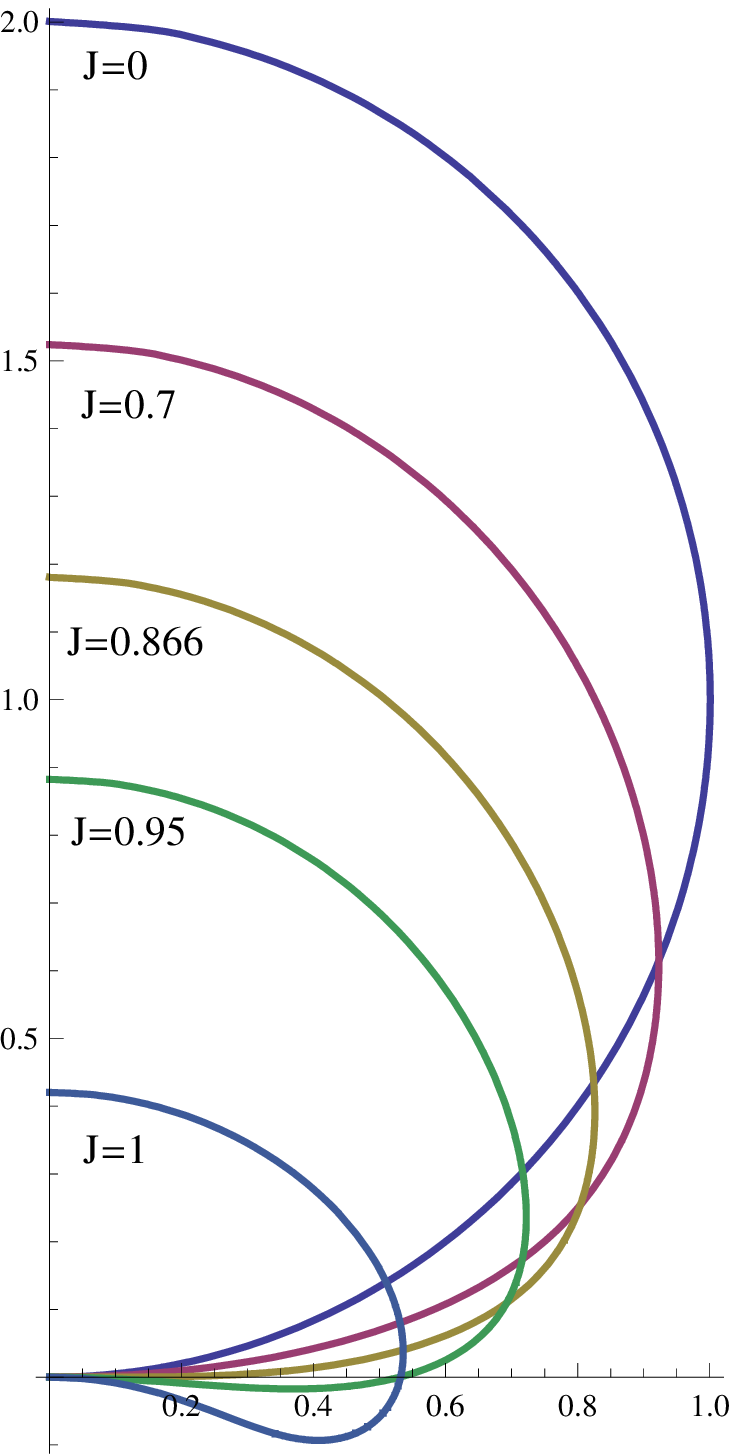}}
\begin{picture}(0,0)(0,0)
\end{picture}
\caption{Profile of the embedding in $\mathbb{E}^3$ (left, first presented in \cite{Costa:2009wj}) and $H^3$ (right, using $k=1$)  of the Kerr horizon, for fixed mass $M=1$ and various values of the angular momentum $J$. For the hyperbolic embedding only the shape is relevant, not the overall scaling, since the latter depends on an arbitrary integration constant.}
\label{profilekerr}
\end{figure}

As a first application of the general formulae of section \ref{generalformulae} let us consider 
the Kerr-Newman event horizon. In this case \cite{Smarr}, 
\bequ
g(u)=\frac{(1-u^2)(1+c^2)}{1+c^2u^2} \ , \qquad c\equiv
\frac{J}{M(M+\sqrt{M^2-J^2/M^2})} \ , \qquad
R^2=2M\left(M+\sqrt{M^2-\frac{J^2}{M^2}}\right) \ ,  \eequ
where $M,J$ are the ADM mass and angular momentum of the black
hole. Note that $0\le c\le 1$ for black hole solutions, with the equalities attained for Schwarzschild and extreme Kerr, respectively. The embedding
in flat space will fail when the function 
\bequ
f(u)\equiv (1-g'(u)^2/4)/g(u) \ , 
\eequ
becomes
negative. This function is plotted in Fig. \ref{kerr}. The figure shows that, for $J>\sqrt{3}/2$, the embedding in $\mathbb{E}^3$ fails in two patches around each of the poles (two polar caps). The function $f(u)$ is bounded from below, and has a minimum given by $f=-1$. Therefore to embed the complete surface in
$H^3$ it suffices to take $L=R$, i.e $k=1$. In Fig. \ref{profilekerr} the profiles of the embeddings in $\mathbb{E}^3$ and $H^3$, $z=z(r)$, are displayed. In the latter case, the overall scaling is irrelevant, since it is determined by an arbitrary integration constant.

Surprisingly, the embedding of the round sphere (the $J=0$ case) in $\mathbb{E}^3$ and $H^3$ are identical. The hyperbolic embedding of the round sphere can be treated analytically. From \eqref{embed} one gets
\bequ
(\ln z)'=\frac{u\pm \sqrt{1+k^2}}{1+k^2-u^2} \ . 
\eequ
The two solutions are interchanged by the reparameterisation
$u\rightarrow -u$. Integrating yields
\bequ
z=\frac{\rm \alpha}{\sqrt{1+k^2}-u} \ , \ \ \ \ {\rm \alpha=constant>0} \ . \eequ
Using \eqref{embed}, one arrives at the surface
equation 
\bequ
r^2+\left(z-\frac{\alpha \sqrt{1+k^2}}{k^2}\right)^2=\frac{\alpha^2}{k^4} \ . \label{hypersphere}\eequ
This is indeed the standard equation of a spherical surface (but note that these are the coordinates of the upper half space model for $H^3$). In Fig \ref{profilekerr} we have taken $k=1=\alpha$ and we have performed a $z$ translation so that the plot intersects the origin.

\begin{figure}[h!]
\centering\includegraphics[height=3in,width=2.5in]{{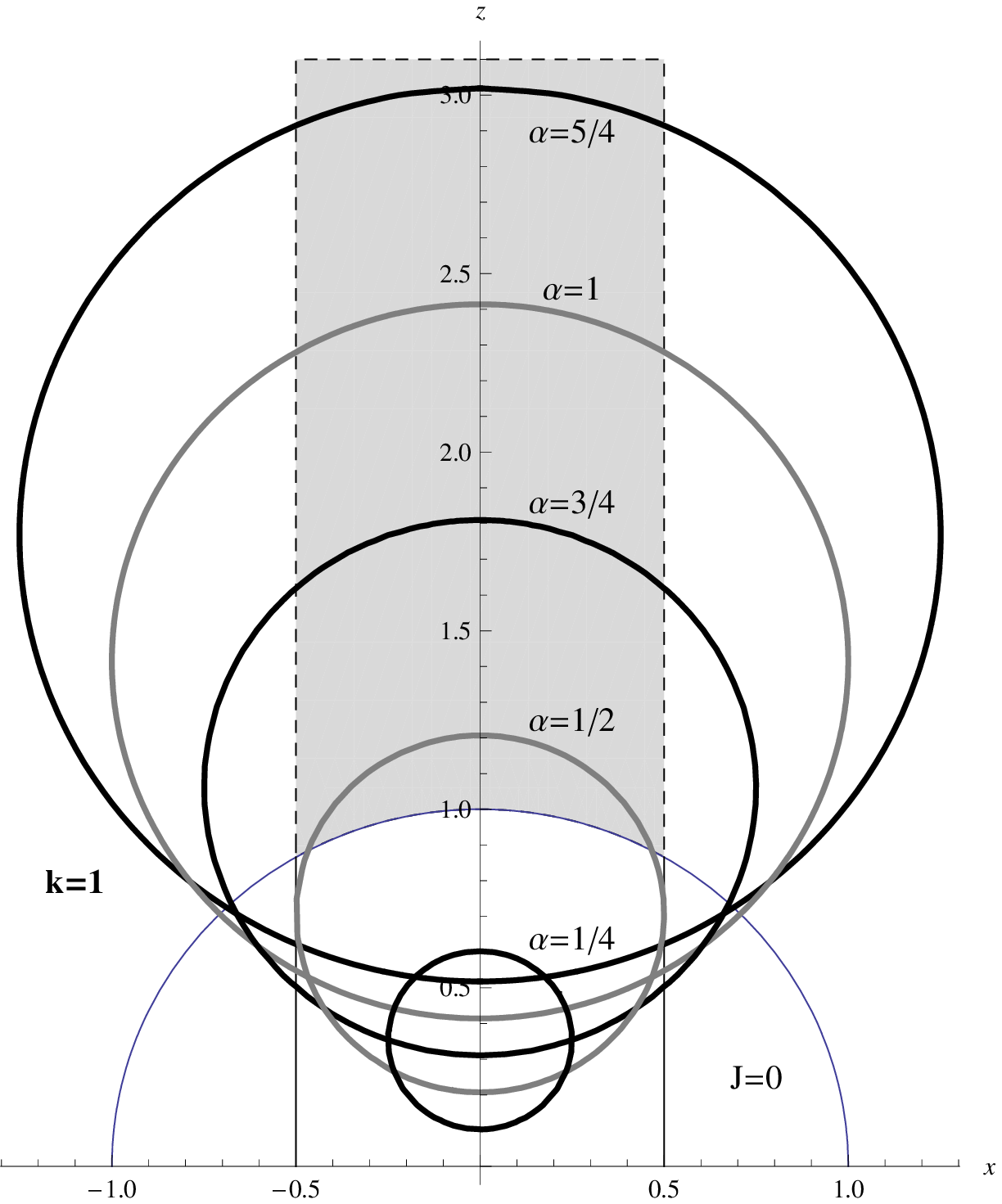}}
\ \ \ \ \ \ \ \ \ \ \ 
\centering\includegraphics[height=3in,width=2.5in]{{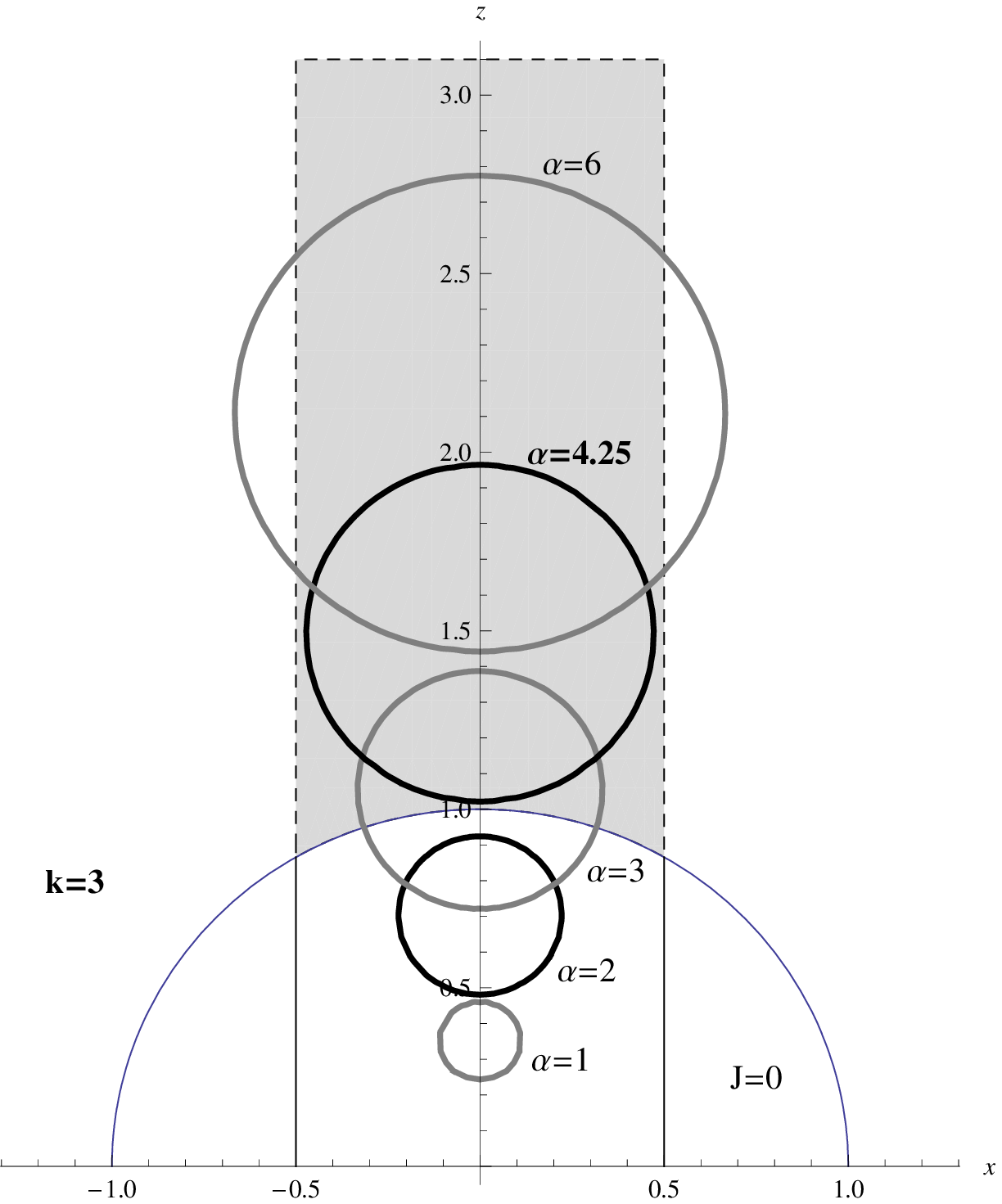}}
\begin{picture}(0,0)(0,0)
\end{picture}
\caption{Embedding the round 2-sphere ($J$=0) into $H^3$ for $k=1$ (left) and $k=3$ (right) and various values of $\alpha$. For $k=1$ ($k=3$) the embedding cannot (can, taking an appropriate $\alpha$) be fitted completely inside the fundamental domain \eqref{fundamentald} - shaded region. We display the $x,z$ plane - as in Fig. \ref{fd11}, \ref{fd2} and \ref{fd21} - but note that the embedded surface has a $U(1)$ isometry in the $x-y$ plane.}
\label{fd1}
\end{figure}

\begin{figure}[h!]
\centering\includegraphics[height=3in,width=2.5in]{{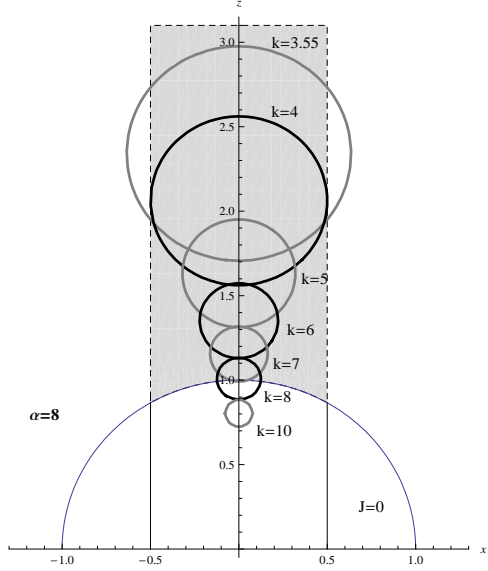}}
\begin{picture}(0,0)(0,0)
\end{picture}
\caption{Embedding the round 2-sphere ($J$=0) into $H^3$ for $\alpha=8$ and various values of $k$. The embedding can, taking the value of $k$ in an appropriate range,  be fitted completely inside the fundamental domain \eqref{fundamentald} - shaded region.}
\label{fd11}
\end{figure}

Observe that the embedding of the round sphere can be fitted in the fundamental domain discussed in section \ref{identifications}. Indeed, if suffices to take $k,\alpha$ such that $\alpha/k^2<1/2$ and $(\sqrt{1+k^2}-1)\alpha/k^2>1$. This is possible if $k$ is sufficiently large and choosing the value of $\alpha$ appropriately - Fig. \ref{fd1}; in this figure the effect of increasing $\alpha$ for constant $k$ can be seen: it raises the centre of the sphere and increases its radius. The same effect is obtained, for constant $\alpha$, \textit{decreasing} the value of $k$ - Fig. \ref{fd11}.

We can use the embedding of $H^3$ in four dimensional Minkowski space given in section \ref{uhsmodel}, to check that \eqref{hypersphere} describes a round sphere. Noting that $r^2/z^2=X^2+Y^2$ and $z=1/(T+Z)$ we find the surface
\bequ
X^2+Y^2+\left(\frac{\alpha}{k}(T+Z)-\frac{\sqrt{1+k^2}}{k}\right)^2=\frac{1}{k^2} \ . \label{elipse} \eequ
The 2-surface described by \eqref{hypersphere} is the intersection of \eqref{elipse} with $H^3$ embedded in ${\Bbb E} ^{3,1}$ as the surface \eqref{hypersurface}. This yields the co-dimension two set in ${\Bbb E} ^{3,1}$ described by 
\bequ
X^2+Y^2+\left(\sqrt{1-v^2}\,Z+\frac{v\sqrt{1+k^2}}{k}\right)^2=\frac{1}{k^2} \ , \ \ \ \ \ \  \frac{1}{\sqrt{1-v^2}}(T+vZ)=\frac{\sqrt{1+k^2}}{k} \ , \label{elipsoid}\eequ
where the ``velocity" $v$ is defined as
\bequ v\equiv \frac{\alpha^2-k^2}{\alpha^2+k^2} \ . \eequ
Introducing ``boosted" coordinates
\bequ
T'=\frac{1}{\sqrt{1-v^2}}(T+vZ) \ , \qquad Z'=\frac{1}{\sqrt{1-v^2}}(Z+vT) \ , \eequ
\eqref{hypersurface} becomes $X^2+Y^2+Z'^2-T'^2=-1$ and the 2-surface  \eqref{elipsoid} becomes
\bequ
X^2+Y^2+Z'^2=\frac{1}{k^2} \ , \ \ \ \ \ \  T'=\frac{\sqrt{1+k^2}}{k}>1 \ . \eequ
This is indeed a round 2-sphere; the apparent ellipsoid exhibited in \eqref{elipsoid} is a result of Lorentz contraction.

\begin{figure}[h!]
\centering\includegraphics[height=2in,width=2in]{{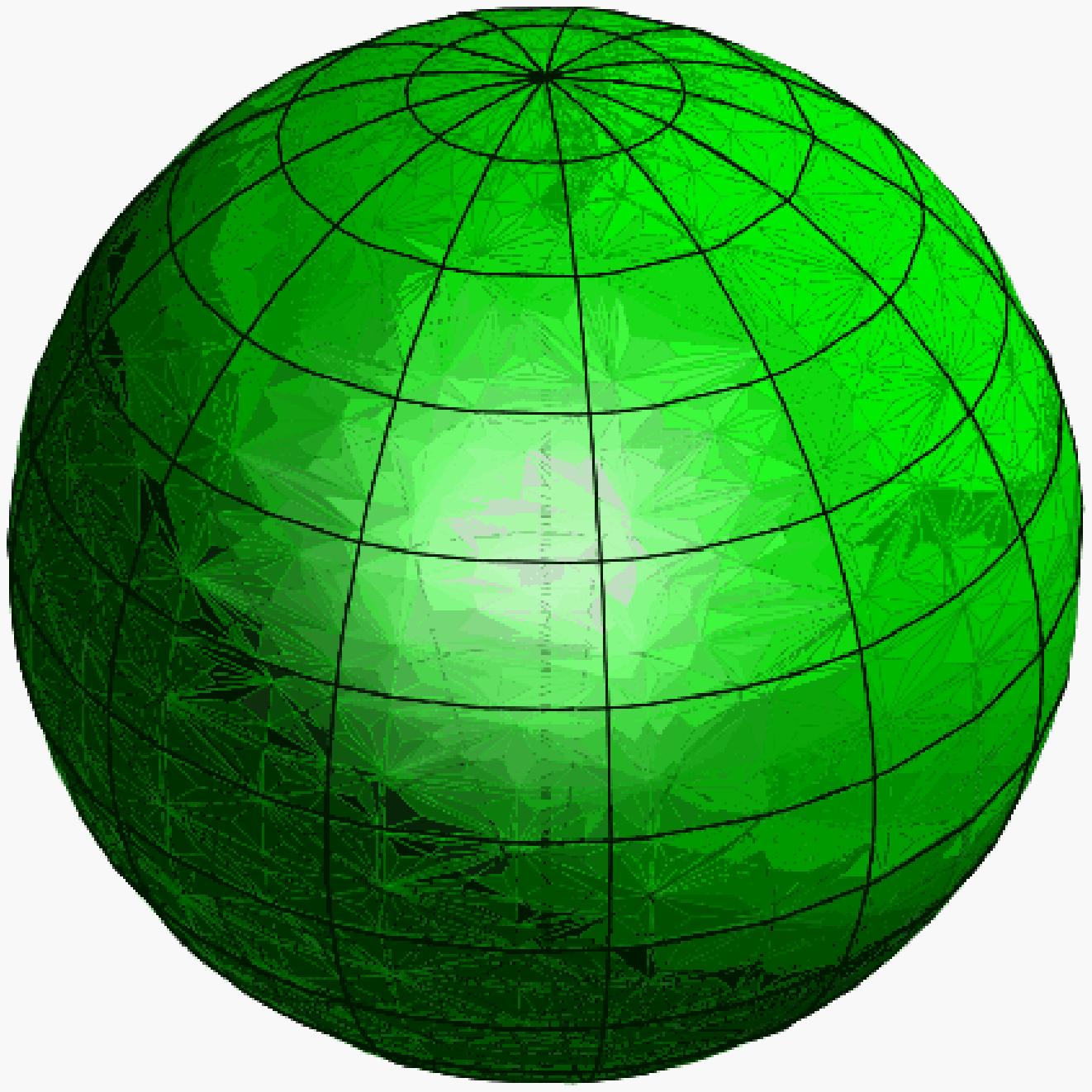}}
\centering\includegraphics[height=2in,width=2in]{{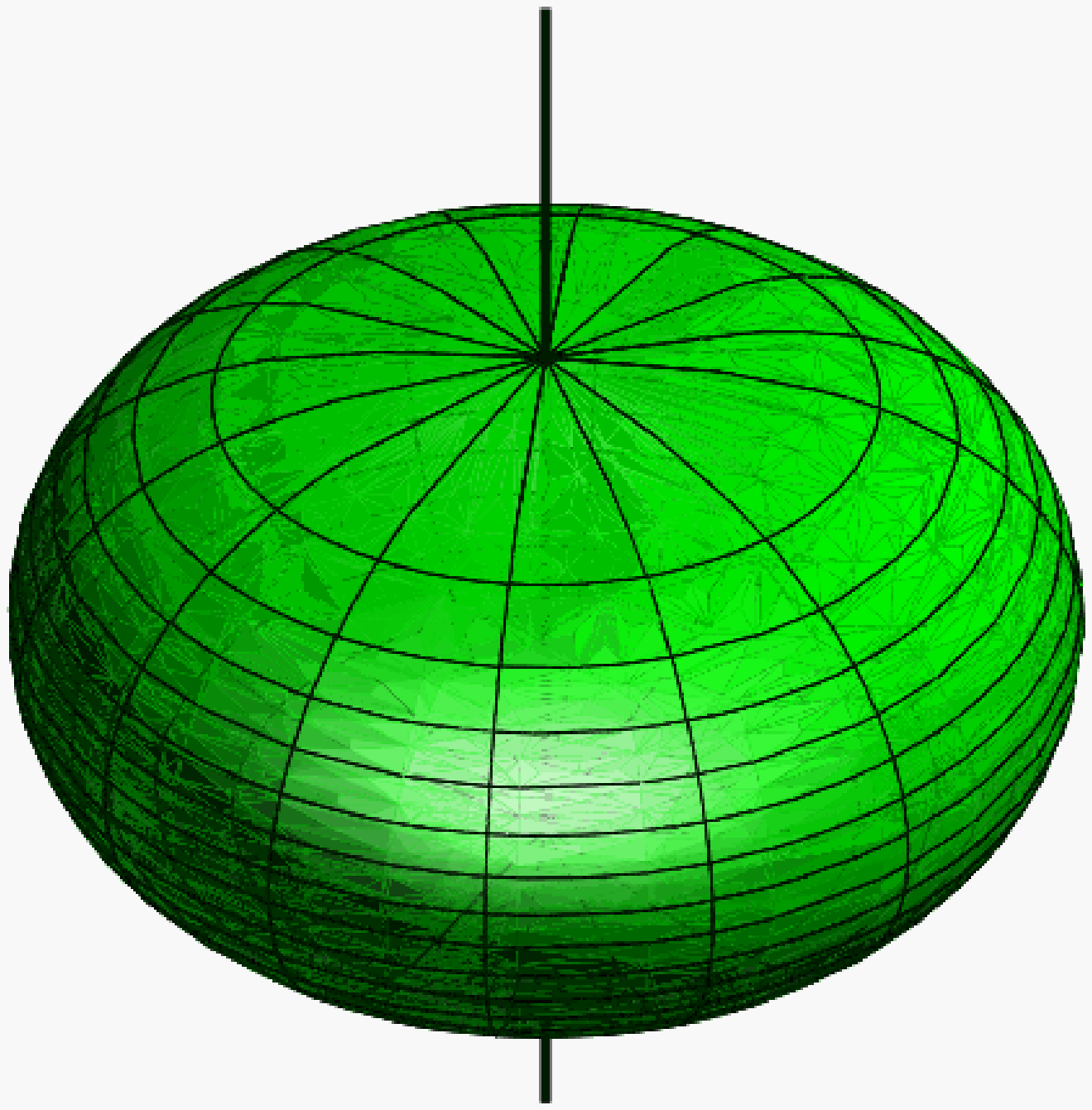}}
\centering\includegraphics[height=2in,width=2in]{{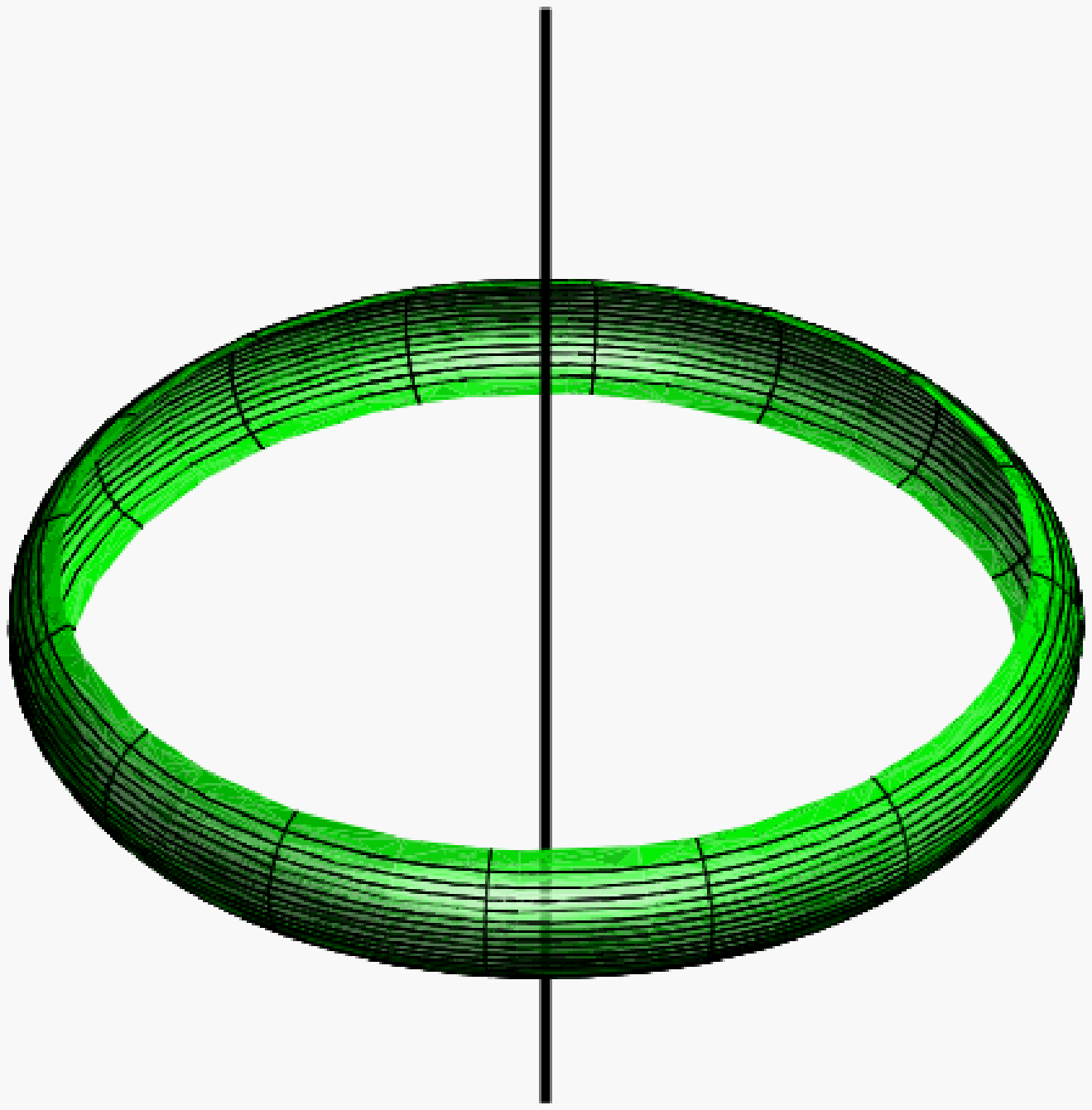}}
\begin{picture}(0,0)(0,0)
\end{picture}
\caption{Embedding the Kerr horizon for fixed mass $M=1$ and various values of the angular momentum $J$ ($J=0,0.86,1$) in $\mathbb{R}^3$. In the extremal case the embedding covers only a region of the horizon around the equator \cite{Smarr}.}
\label{3de}
\end{figure}

\begin{figure}[h!]
\centering\includegraphics[height=2in,width=2in]{{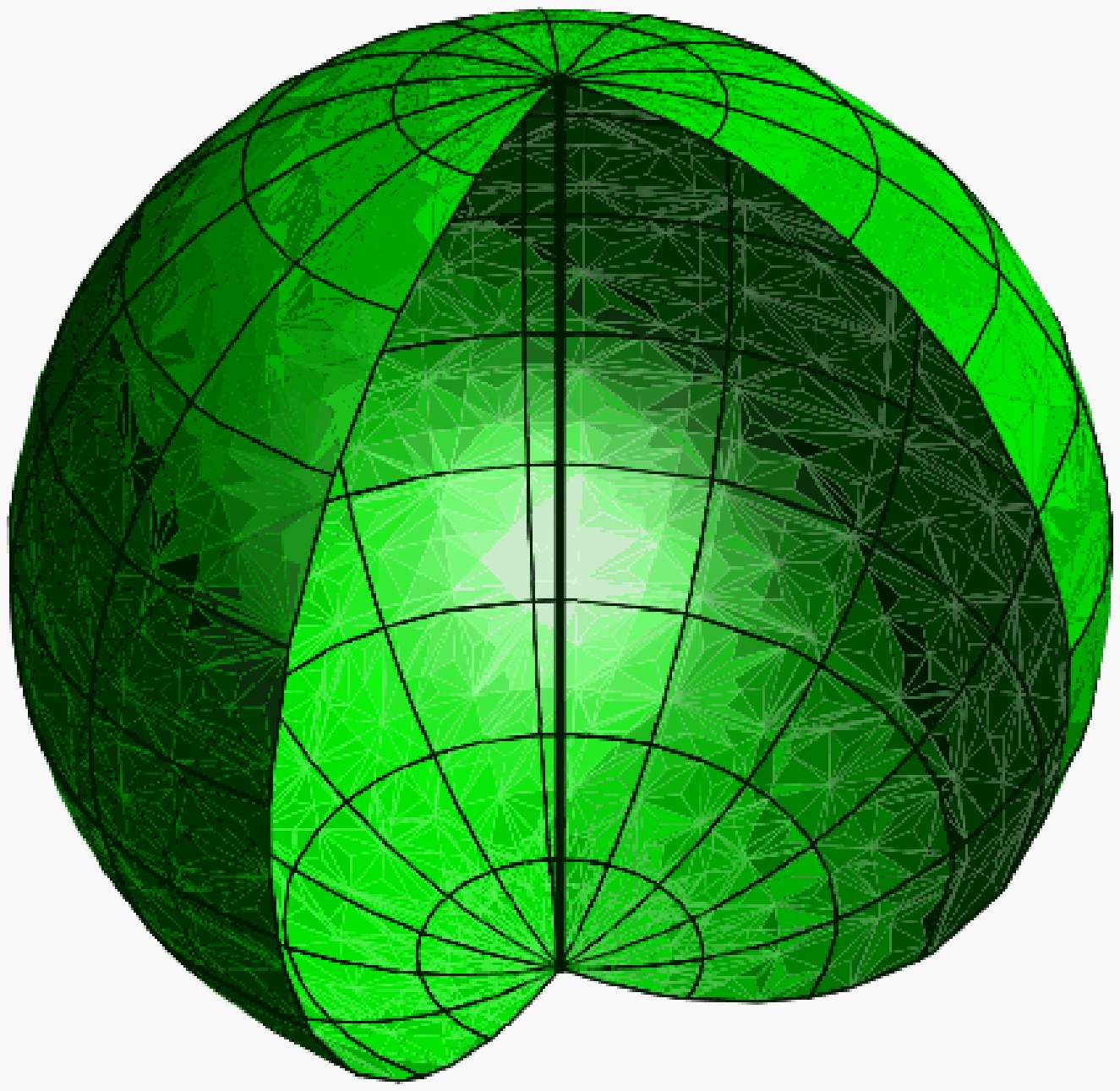}}
\centering\includegraphics[height=2in,width=2in]{{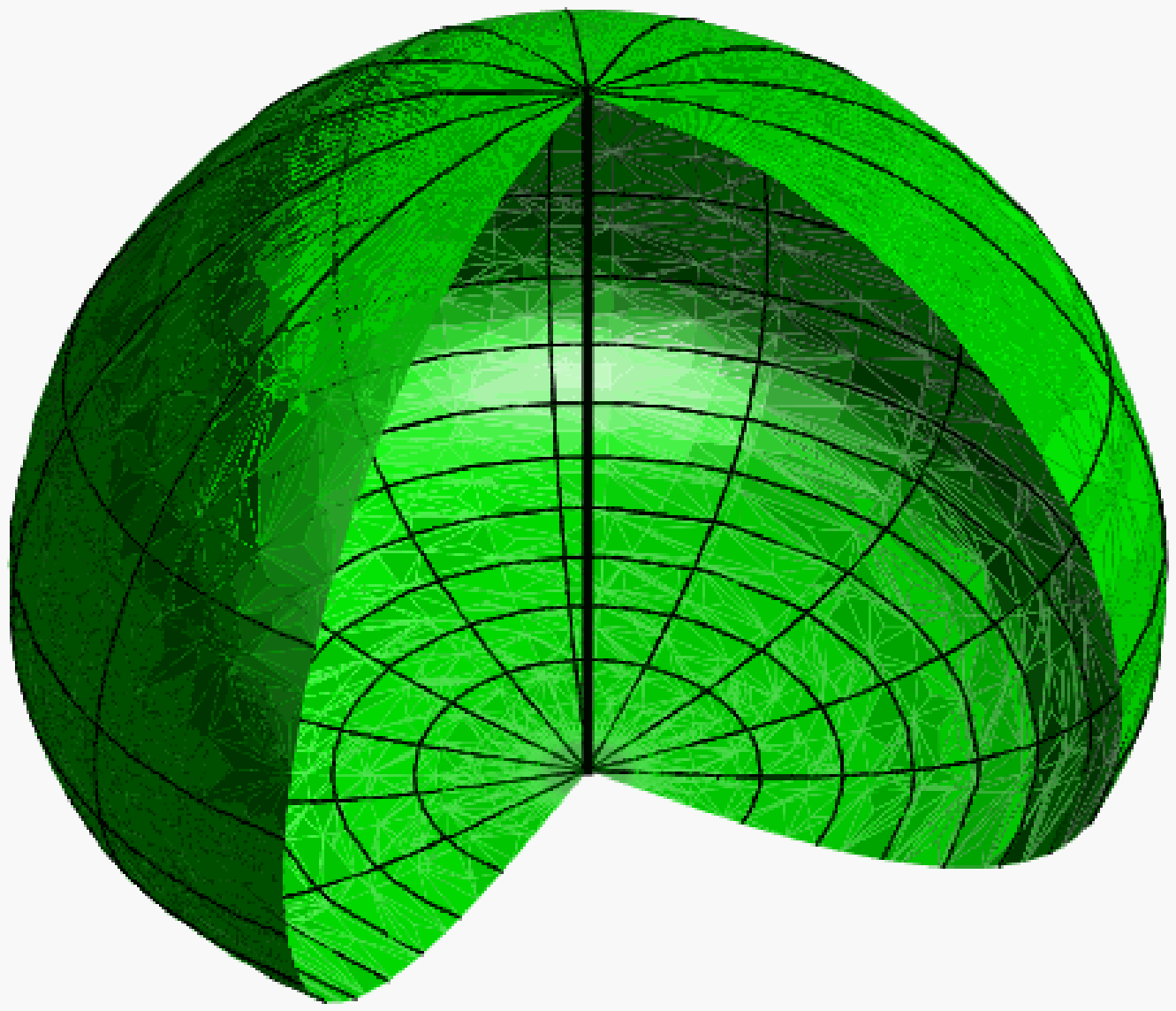}}
\centering\includegraphics[height=2in,width=2in]{{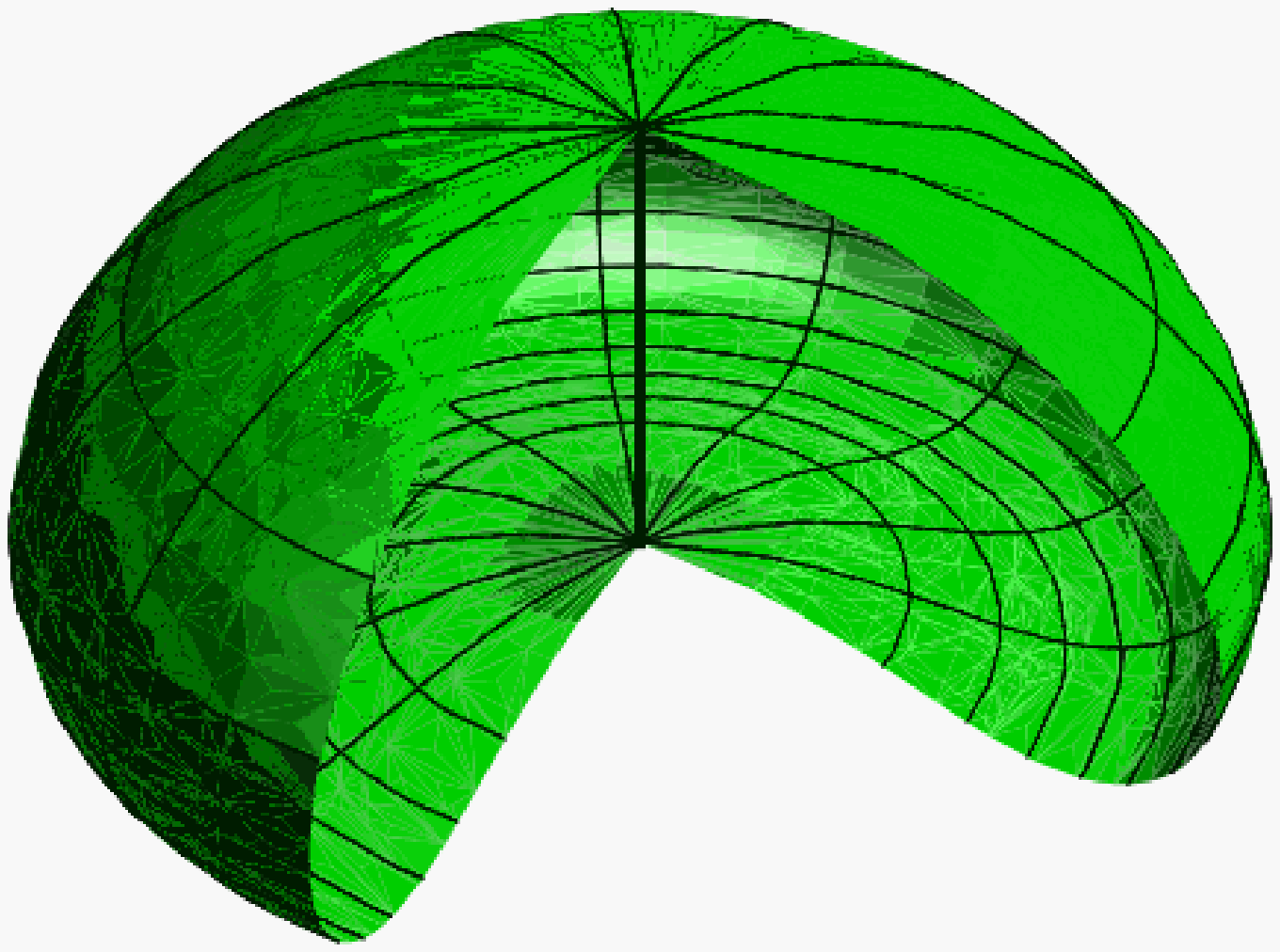}}
\begin{picture}(0,0)(0,0)
\end{picture}
\caption{Embedding the Kerr horizon for fixed mass $M=1$ and various values of the angular momentum $J$  ($J=0,0.86,1$) in $H^3$, with $k=1$. The embedding covers the whole of the horizon, even in the extremal case. Note, however, that the $\mathbb{Z}_2$ symmetry around the equator is lost, when the angular momentum is turned on. This is to be expected, since it is not a symmetry of the embedding space. But surprisingly the symmetry remains in the $J=0$ case.}
\label{3dh}
\end{figure}

In Fig. \ref{3de} and Fig. \ref{3dh} we present 3D plots of the embeddings of the Kerr-Newman event horizon in $\mathbb{E}^3$ and $H^3$, respectively, for various values of the angular momentum. In the hyperbolic case the embedding is global for any value of $J$. Note that in the Euclidean case, even a local embedding of the region around the poles is impossible for $J> \sqrt{3}/2$ \cite{Frolov}. 

\begin{figure}[h!]
\centering\includegraphics[height=3in,width=3in]{{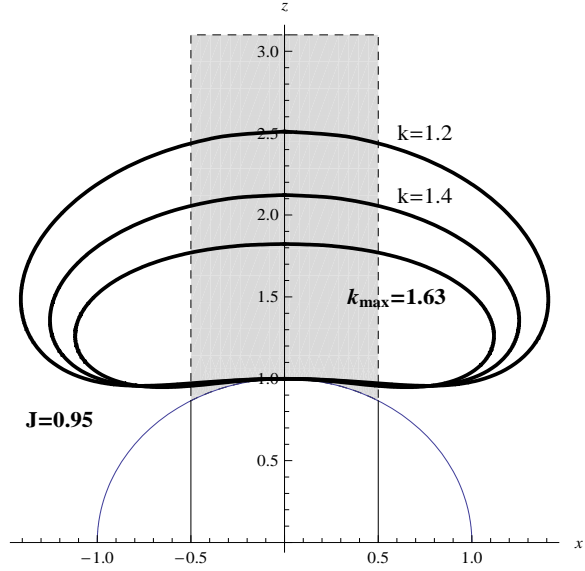}}
\begin{picture}(0,0)(0,0)
\end{picture}
\caption{Embedding the Kerr-Newman event horizon for $M=1$ and $J=0.95$ into $H^3$ for various values of $k$ and of an additional integration constant $\alpha$, fixed at $\alpha=\alpha_{touch}$. No choice can fit the embedding completely inside the fundamental domain \eqref{fundamentald} (shaded region).}
\label{fd2}
\end{figure}

\begin{figure}[h!]
\centering\includegraphics[height=2.6in,width=2.6in]{{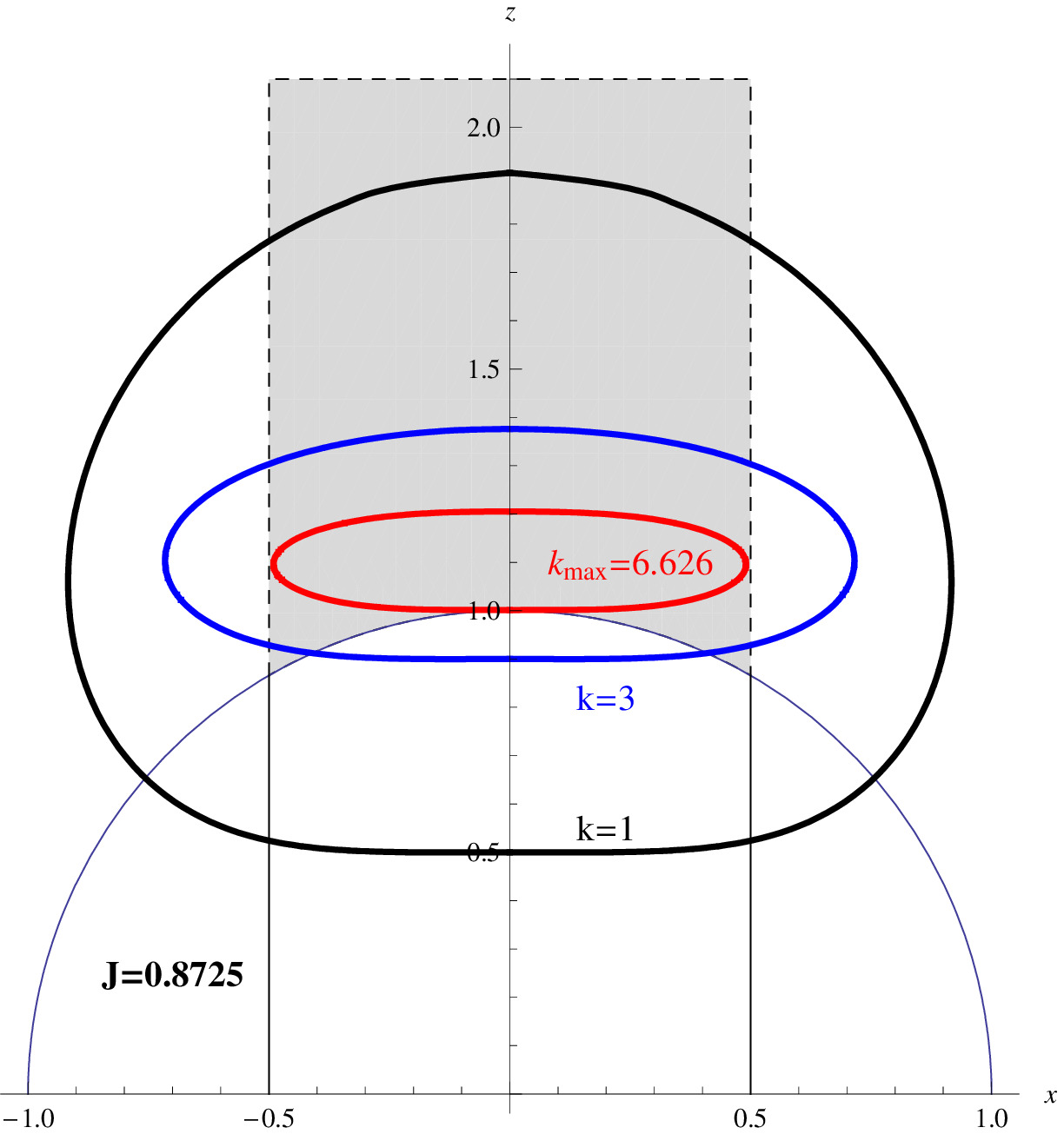}}
\centering\includegraphics[height=3.0in,width=1.0in]{{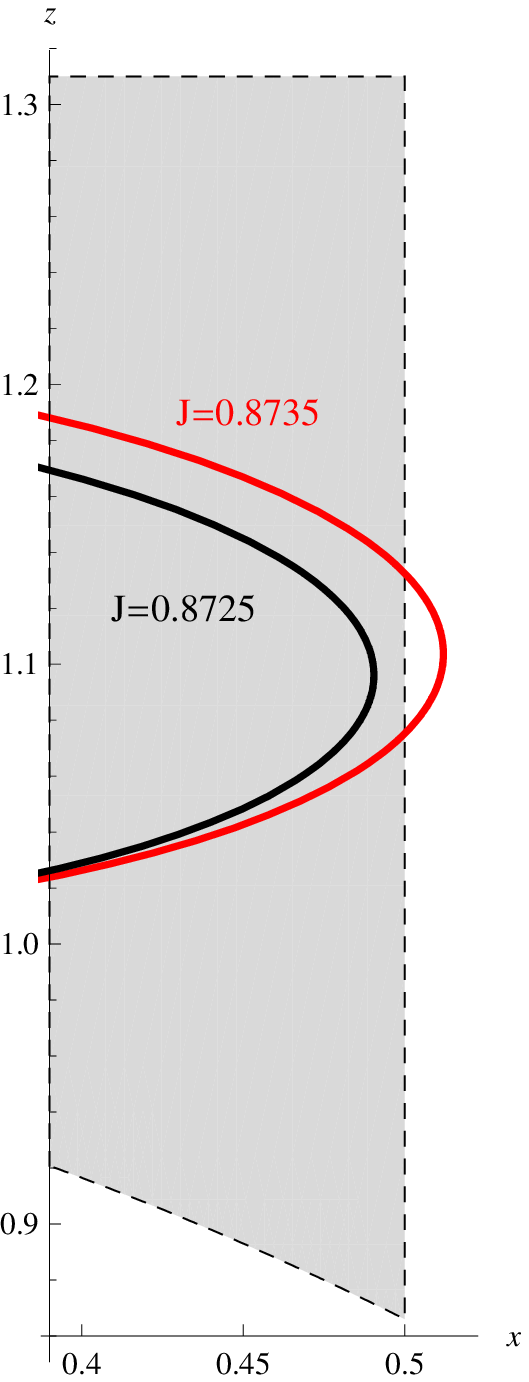}}
\centering\includegraphics[height=2.6in,width=2.6in]{{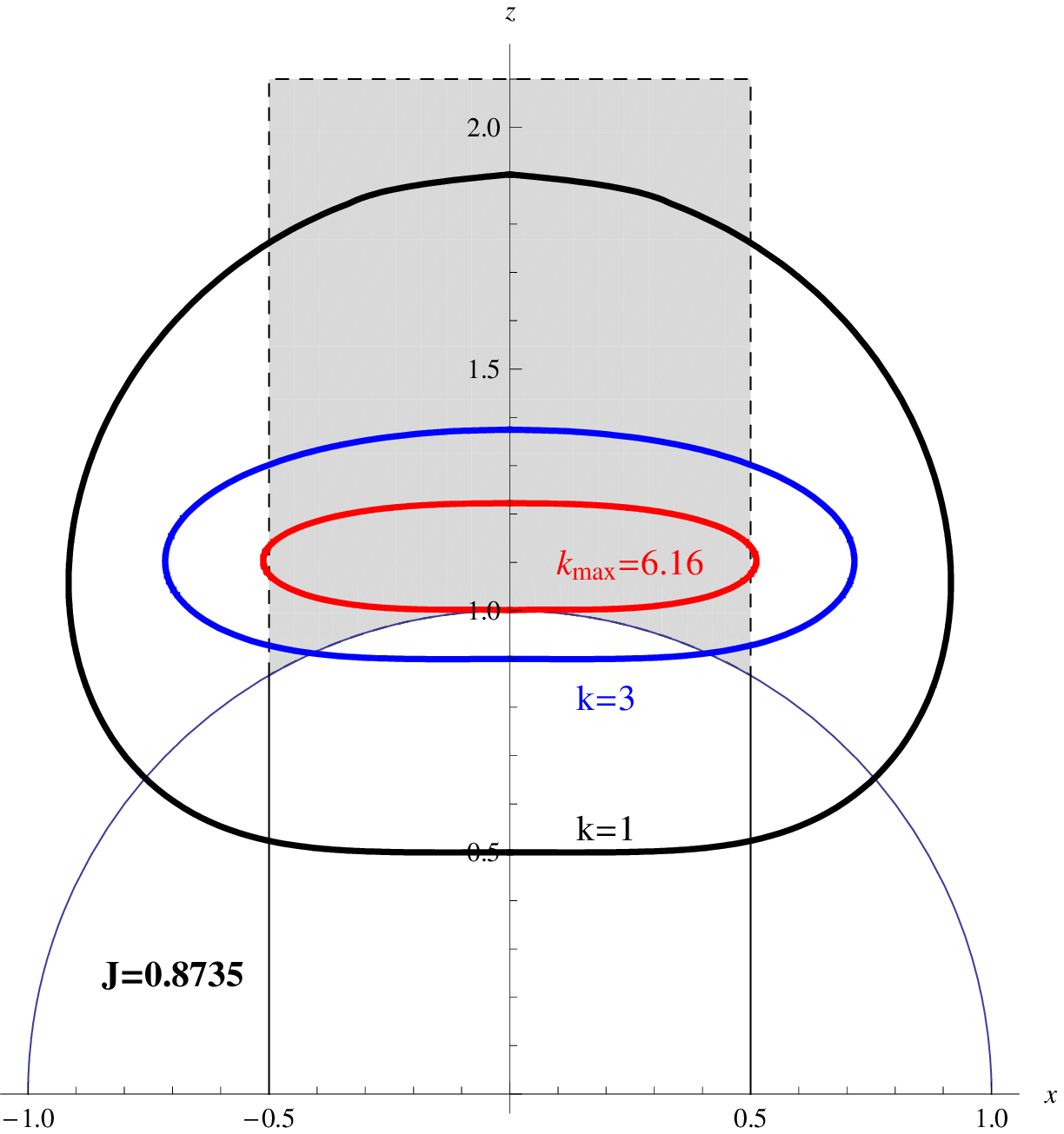}}
\begin{picture}(0,0)(0,0)
\end{picture}
\caption{Embedding the Kerr-Newman event horizon for $M=1$, $J=0.8725$ (left) and $J=0.8735$ (right) into $H^3$ for various values of $k$, including for each case $k_{max}$ and $\alpha=\alpha_{touch}$. In the former case the embedding can still be fitted inside the fundamental domain \eqref{fundamentald} (shaded region), but not in the latter case, which is explicit in the detail (middle).}
\label{fd21}
\end{figure}

For $J\neq 0$ it is still possible to fit the embedding completely inside the fundamental domain, but only up to a maximal value of $J/M^2$ which is in $[0.8725,0.8735]$. This interval has been determined numerically with the following strategy. First observe that, fixing the black hole, i.e $J$ and $M$, the embedding depends on $k$ and on an additional integration constant $\alpha$, as in the $J=0$ case treated above analytically.  Fixing $k$, the effect of increasing $\alpha$ is to move the embedding profile up in the $z$ coordinate, making it simultaneously larger, as in the $J=0$ case (cf. Fig. \ref{fd1}). Thus, our strategy is:
\begin{itemize}
\item[i)] for each $k$, $\alpha$ is fixed at $\alpha=\alpha_{touch}$, such that the points in the embedding profile with the smallest $z$ touch tangentially the boundary of the fundamental domain at $r^2+z^2=1$ ;
\item[ii)] analysing different values of $k$, each with $\alpha=\alpha_{touch}$, one realises that the effect of decreasing $k$ is to increase the size of the profile - see Fig. \ref{fd2} for an example. 
\item[iii)] since $k$ is bounded above for $J/M^2=\sqrt{3}/2$ we investigate, beyond this value of the angular momentum, the embedding profile for the maximum allowed $k=k_{max}$ and $\alpha=\alpha_{touch}$.
\end{itemize} 
It turns out that for values of $J/M^2$ slightly above $\sqrt{3}/2\simeq 0.866$ (we checked up to $0.8725$) it is still possible to fit the embedding in the fundamental domain; but for $J/M^2=0.8735$ this is not the case anymore - Fig. \ref{fd21}.

\section{The double-Kerr horizon}
As a second example we shall now consider the embedding of the double-Kerr event horizon. The double-Kerr solution is a 7-parameter vacuum spacetime. It was originally generated using a B\"acklund transformation \cite{Neugebauer:1980}. But it can also be generated using the inverse scattering technique \cite{Herdeiro:2008kq}. The general solution is extremely complex. But two 3-parameter sub-families have been recently analysed. These are the counter rotating \cite{Herdeiro:2008kq} and co-rotating case \cite{Costa:2009wj}. These two families describe two stationary co-axial Kerr black holes, with the same mass and the same (co-rotating) or opposite (counter-rotating) angular momentum. They are asymptotically flat and they obey the ``axis condition", which guarantees that the azimuthal Killing vector field has zero norm on the symmetry axis. The three parameters of the solutions are, therefore, the mass of each black hole ($M_1=M_2\equiv M$), their angular momentum ($J_1=\pm J
 _2\equiv J$) and the distance between them $\zeta$. The latter parameter is the coordinate 
 distance in Weyl canonical coordinates. But it is a measure of the proper distance $d$, in the sense that the latter is a monotonic function of $\zeta$ \cite{Costa:2009wj}.

The asymptotically flat double Kerr solution for two black holes always has a strut. Physically, this strut provides the necessary force to keep the two black holes in equilibrium. Mathematically it is described by a conical singularity with a conical excess.

\begin{figure}[h!]
\centering\includegraphics[height=2in,width=3in]{{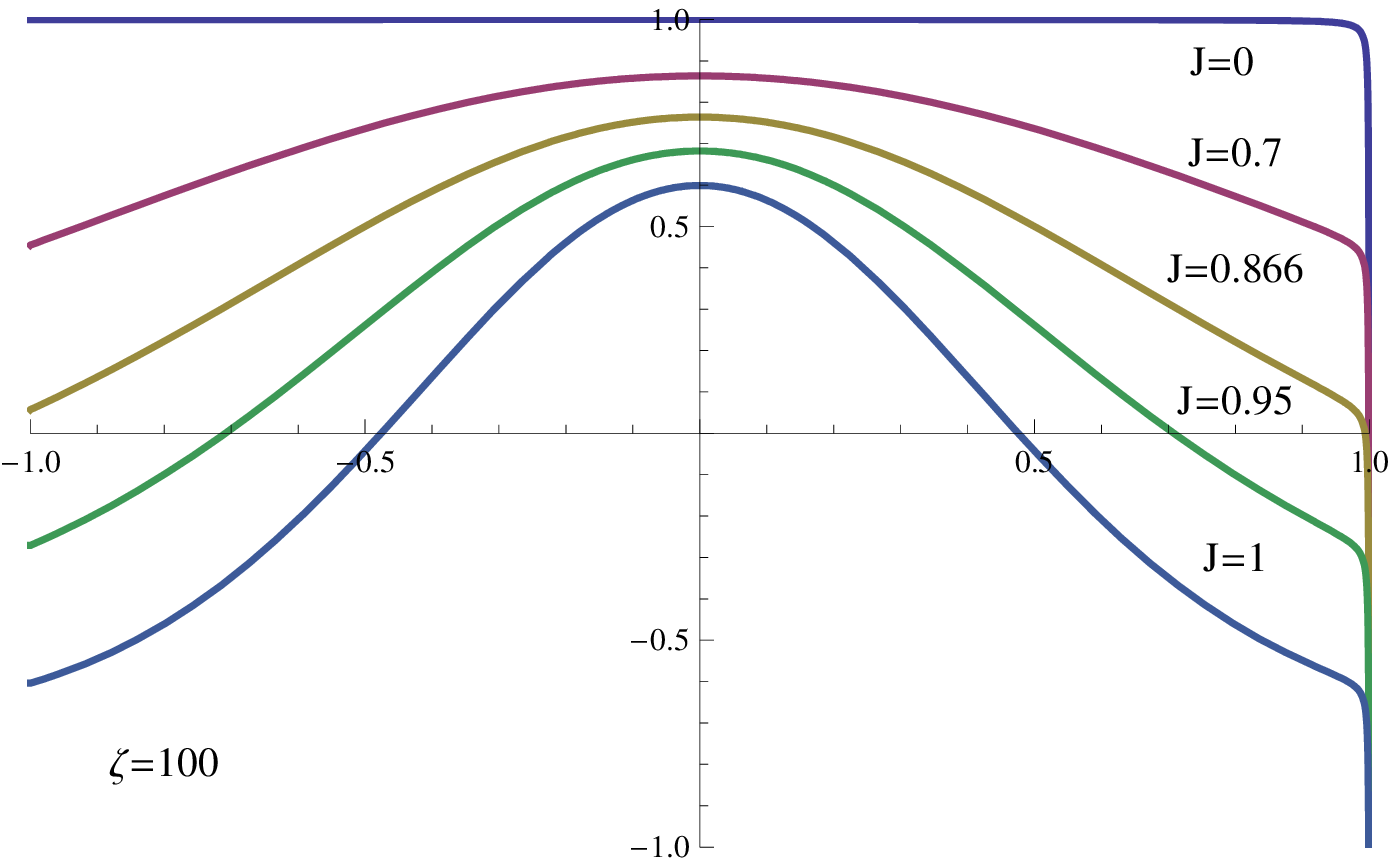}} \ \ \ \ \ \ \ 
\centering\includegraphics[height=2in,width=3in]{{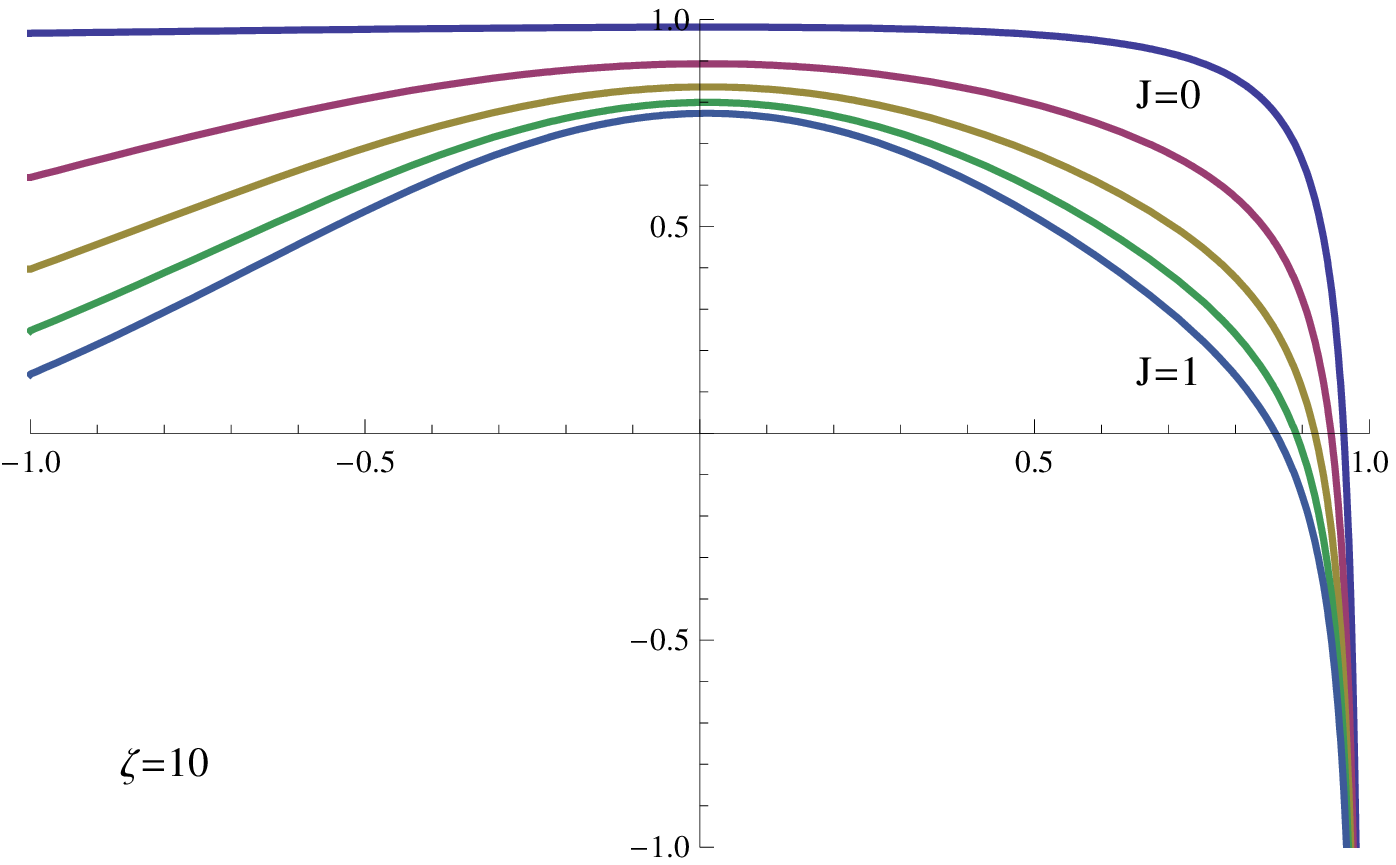}}
\begin{picture}(0,0)(0,0)
\end{picture}
\caption{Function $f(u)$ for $\zeta=100$ (left) and $\zeta=10$ (right), fixed mass $M=1$ and various values of the angular momentum $J$. The left plot corresponds to a large distance and therefore resembles Fig. \ref{kerr} (i.e the case of an isolated black hole), except at the north pole ($u=1$), where the strut meets the horizon. }
\label{doublekerr1}
\end{figure}

\begin{figure}[h!]
\centering\includegraphics[height=2in,width=3in]{{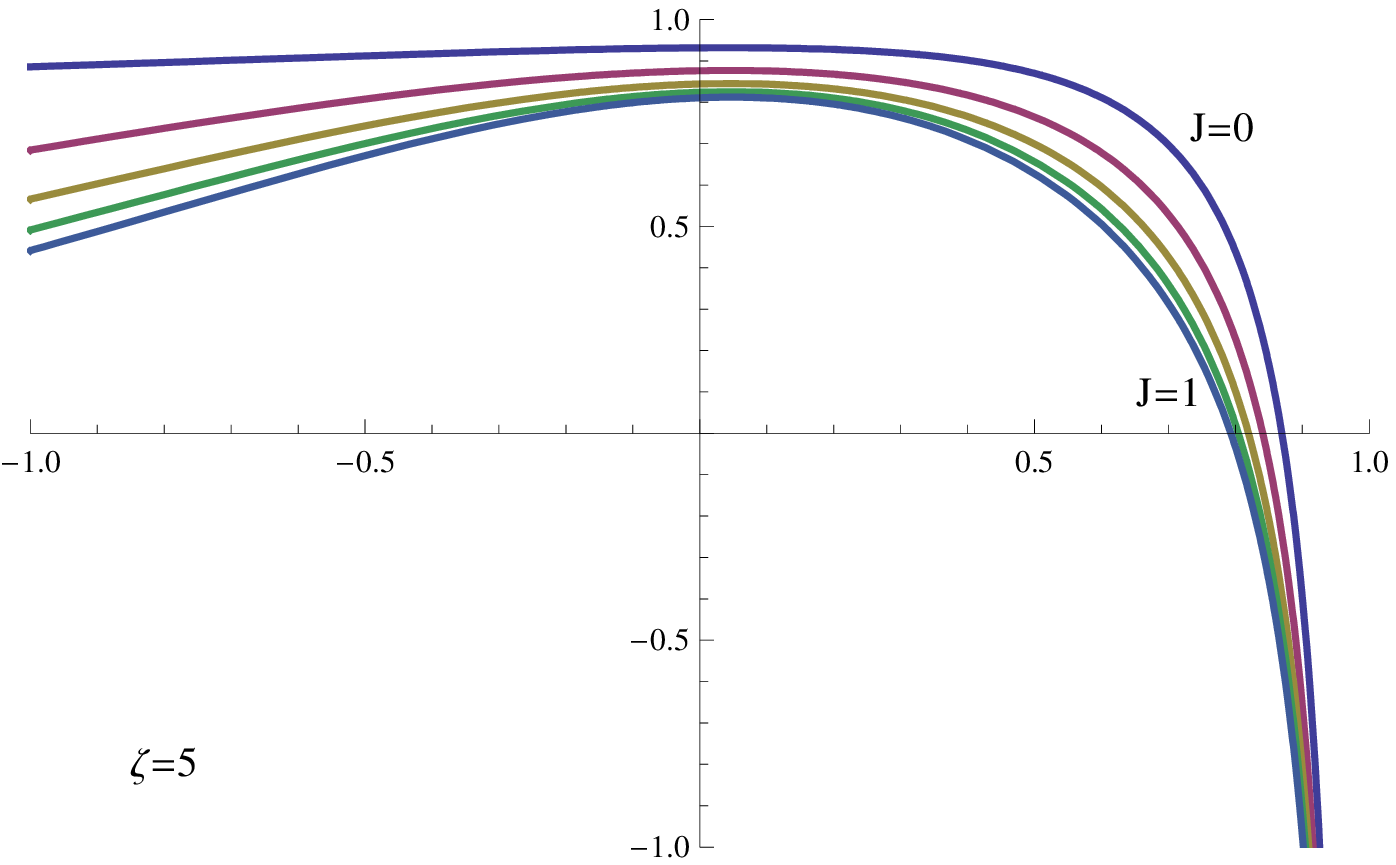}}\ \ \ \ \ \ \ 
\centering\includegraphics[height=2in,width=3in]{{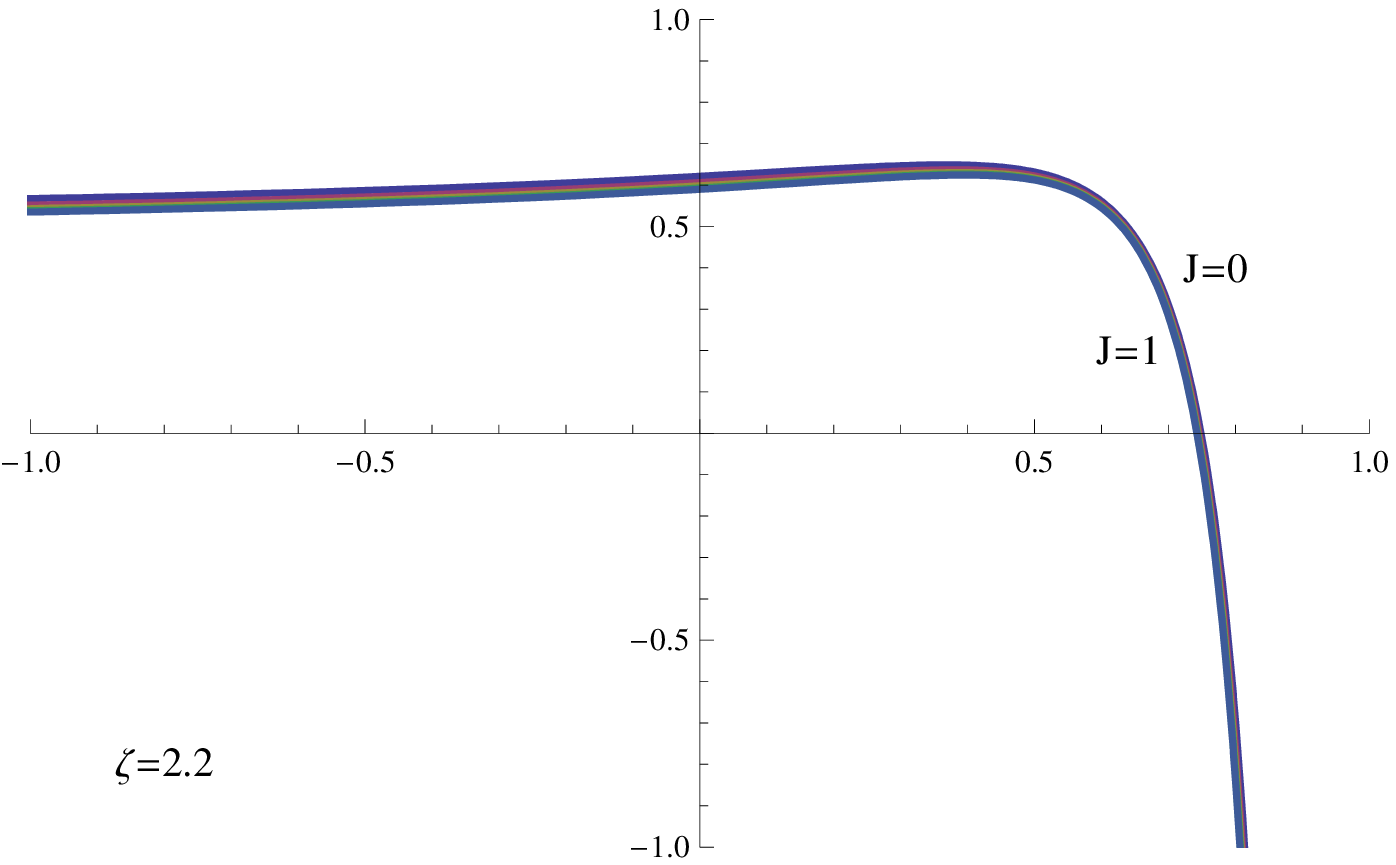}}
\begin{picture}(0,0)(0,0)
\end{picture}
\caption{Function $f(u)$ for $\zeta=5$ (left) and $\zeta=2.2$ (right), fixed mass $M=1$ and various values of the angular momentum $J$. As the distance decreases, the function becomes less sensitive to the angular momentum, and approaches that of a non-rotating black hole, albeit deformed near the north pole.}
\label{doublekerr2}
\end{figure}

In \cite{Costa:2009wj}, the embedding of the double-Kerr horizon was performed in $\mathbb{E}^3$. We shall now perform it in $H^3$. The first observation is that, unlike the Kerr horizon, the embedding in hyperbolic space is not global. The reason can be appreciated in Fig. \ref{doublekerr1} and  \ref{doublekerr2}, where we have plotted the function $f(u)$, for the counter-rotating case, for various values of $\zeta$ and $J$, fixing $M=1$ and choosing the ``lower" black hole in the double-Kerr system, i.e the one with the strut on its north pole ($u=1$). It can be seen that the function $f(u)$ always diverges at $u=1$. This is to be expected and corresponds to the location of the strut. Thus, irrespectively of the value of $k$, the embedding will never completely cover the surface.

Fig. \ref{doublekerr1} shows that the behaviour of $f(u)$ is very similar in the double Kerr system and in the single Kerr system for large distance ($\zeta=100$). There is however, a difference in behaviour at the north pole, where the strut meets the horizon. As the distance between the two black holes decreases, the function $f(u)$ approaches that of a non-rotating black hole, albeit somewhat deformed, cf. Fig. \ref{doublekerr1} and  \ref{doublekerr2}. This was interpreted in \cite{Herdeiro:2008kq} as being caused by the mutual rotational dragging slow down that the black holes exert on one another. 

\begin{figure}[h!]
\centering\includegraphics[height=2.5in,width=2.5in]{{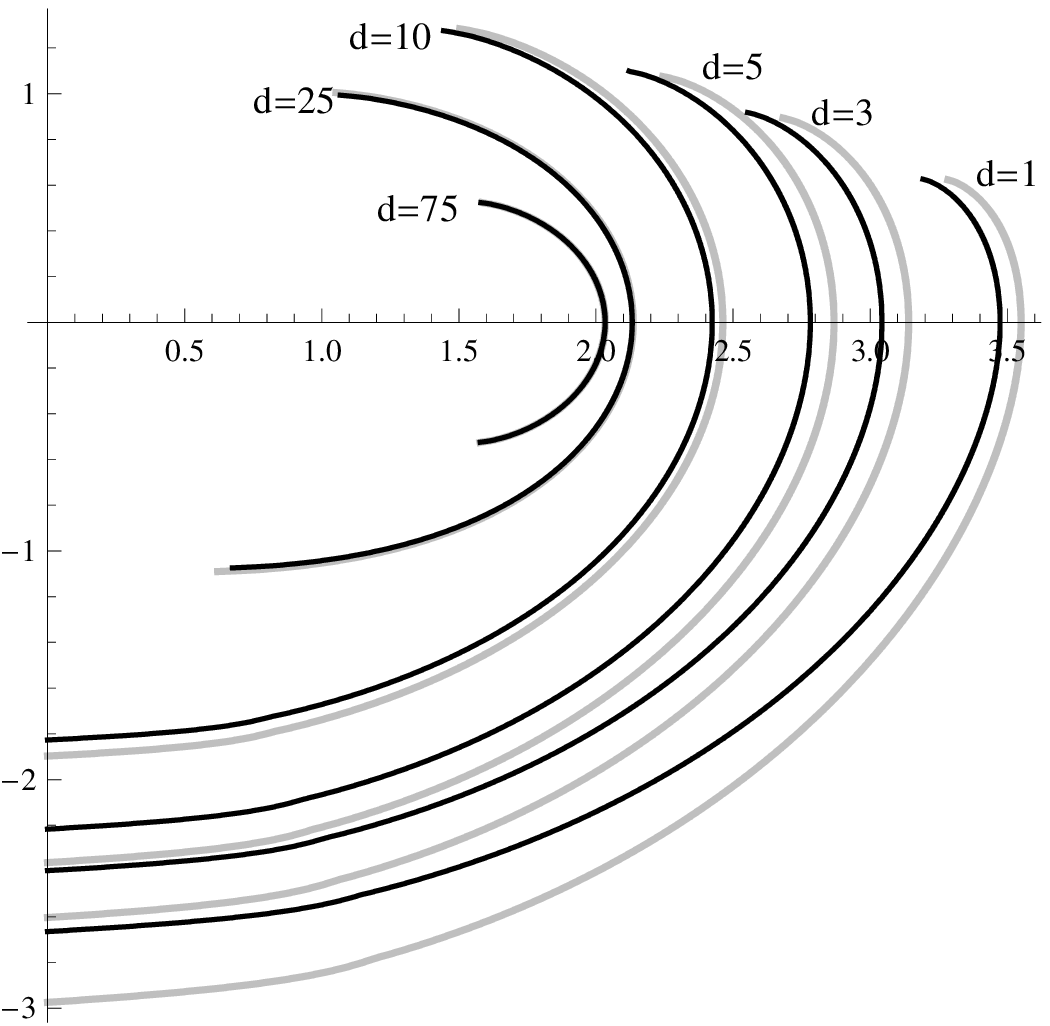}}
\ \ \ \ \ \ \ \ \ \ \ \ 
\centering\includegraphics[height=2.5in,width=2.5in]{{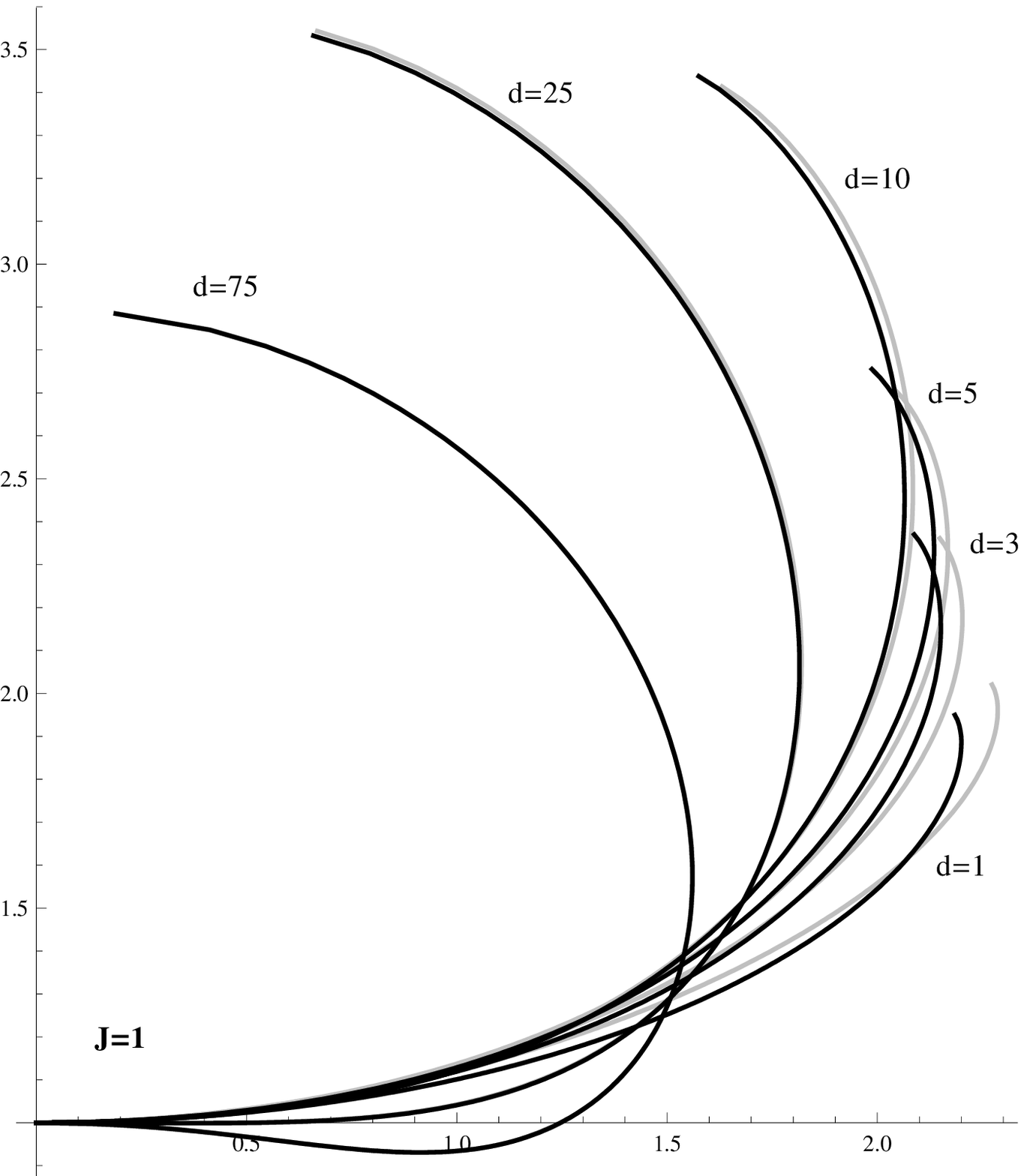}}
\begin{picture}(0,0)(0,0)
\end{picture}
\caption{Profile of the embedding in $\mathbb{E}^3$ (left, first presented in \cite{Costa:2009wj}) and $H^3$ (right, using $k=1$)  of the double-Kerr horizon, for fixed mass and angular momentum $M=1=J$ and various values of the \textit{physical} distance in between the two black holes. At the north pole there is a "strut", and the embedding always fails. The grey (black) lines correspond to the counter-rotating (co-rotating) case.}
\label{profile}
\end{figure}

One may ask how much of the horizon surface is covered by the hyperbolic embedding, fixing a value of $k$. Somewhat surprisingly, choosing the minimal value of $k$ that covers the whole surface in the limit of infinite distance ($k=1$) and for two black holes with $|J|=M^2$, the hyperbolic embedding does not appear to cover more of the horizon surface than the Euclidean one, for small distances - Fig. \ref{profile}.

\begin{figure}[h!]
\centering\includegraphics[height=2in,width=2in]{{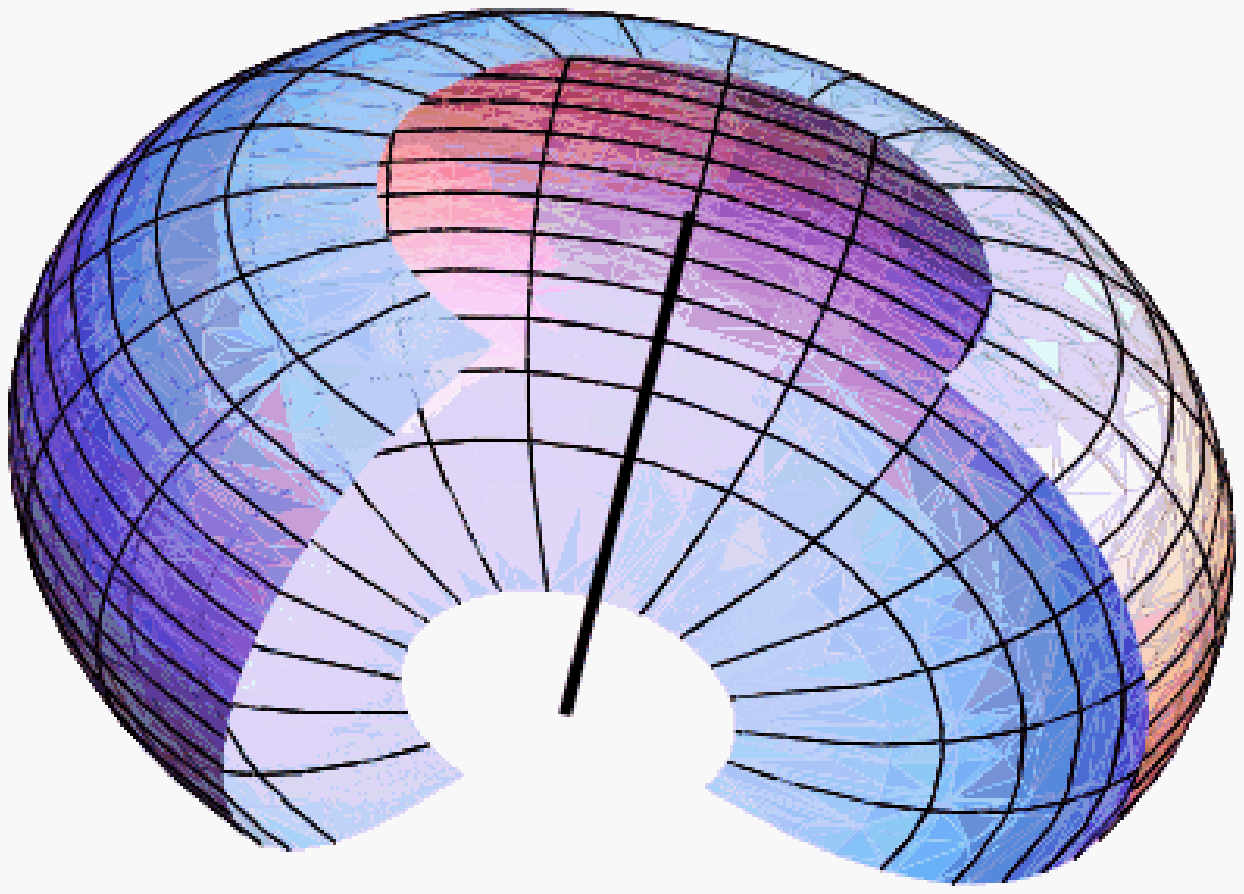}}
\centering\includegraphics[height=2in,width=2in]{{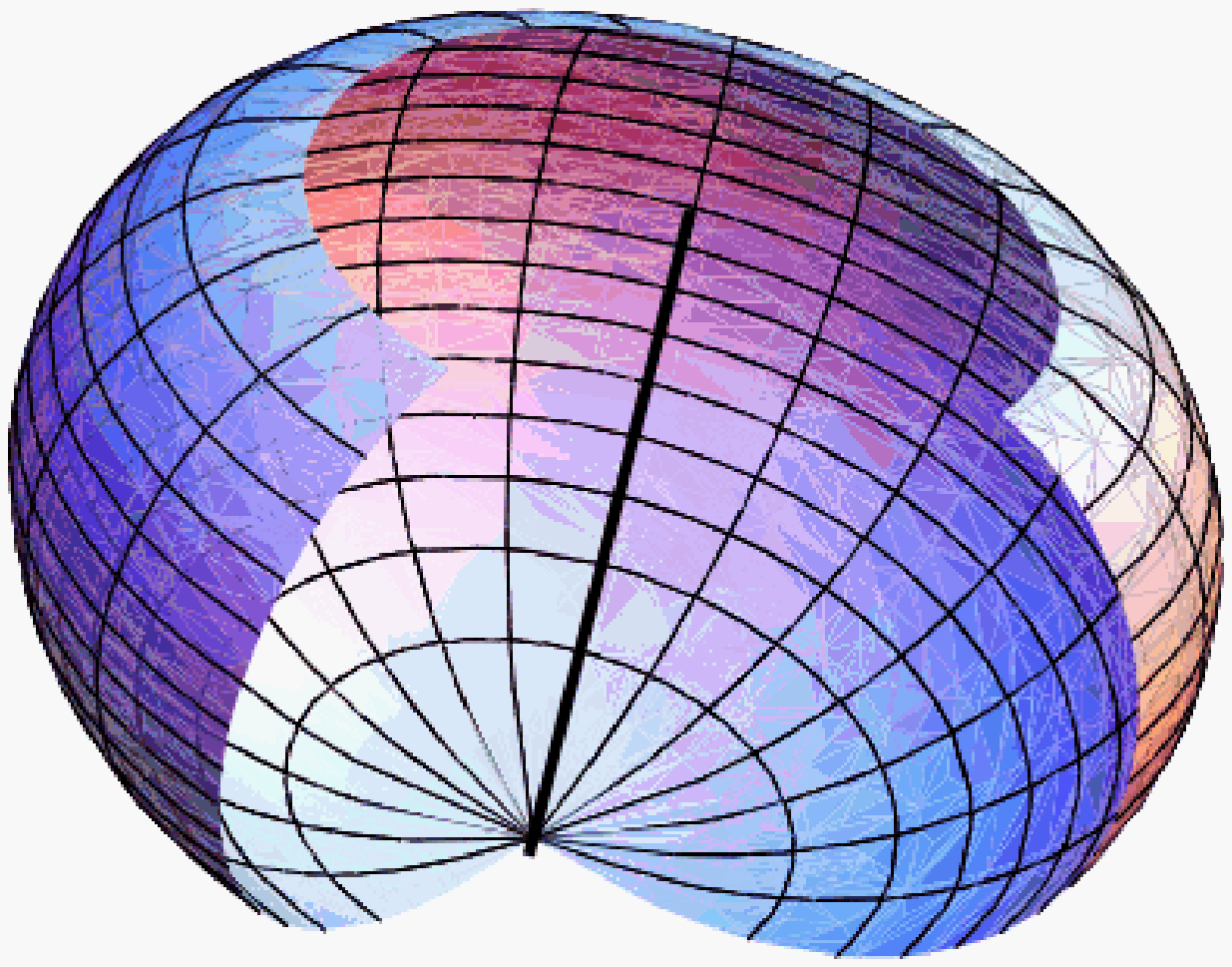}}
\centering\includegraphics[height=2in,width=2in]{{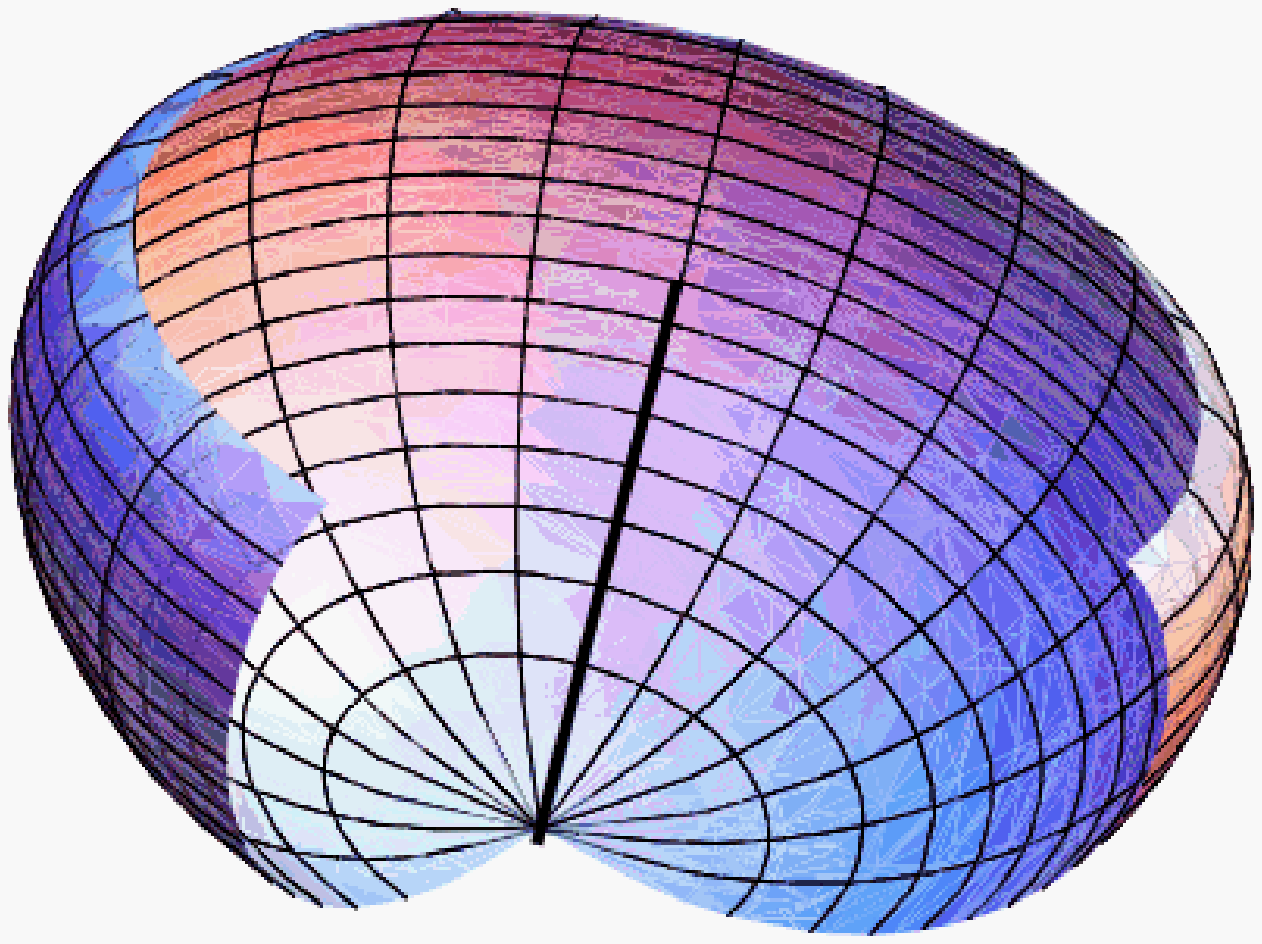}}
\begin{picture}(0,0)(0,0)
\end{picture}
\caption{Embedding the ``lower" black hole in double Kerr horizon for fixed mass and angular momentum $M=1=J$ and various values of the physical distance $d$ ($d=25,10,3$) in $\mathbb{E}^3$. As the distance decreases: i) the south pole gets covered by the embedding (this is a consequence of the smaller angular velocity of the black hole); ii) there is a growing patch around the north pole that is not covered by the embedding, manifesting the deformation induced by the strut, which becomes stronger as the interaction between the two black holes becomes stronger.}
\label{dke}
\end{figure}

\begin{figure}[h!]
\centering\includegraphics[height=2in,width=2in]{{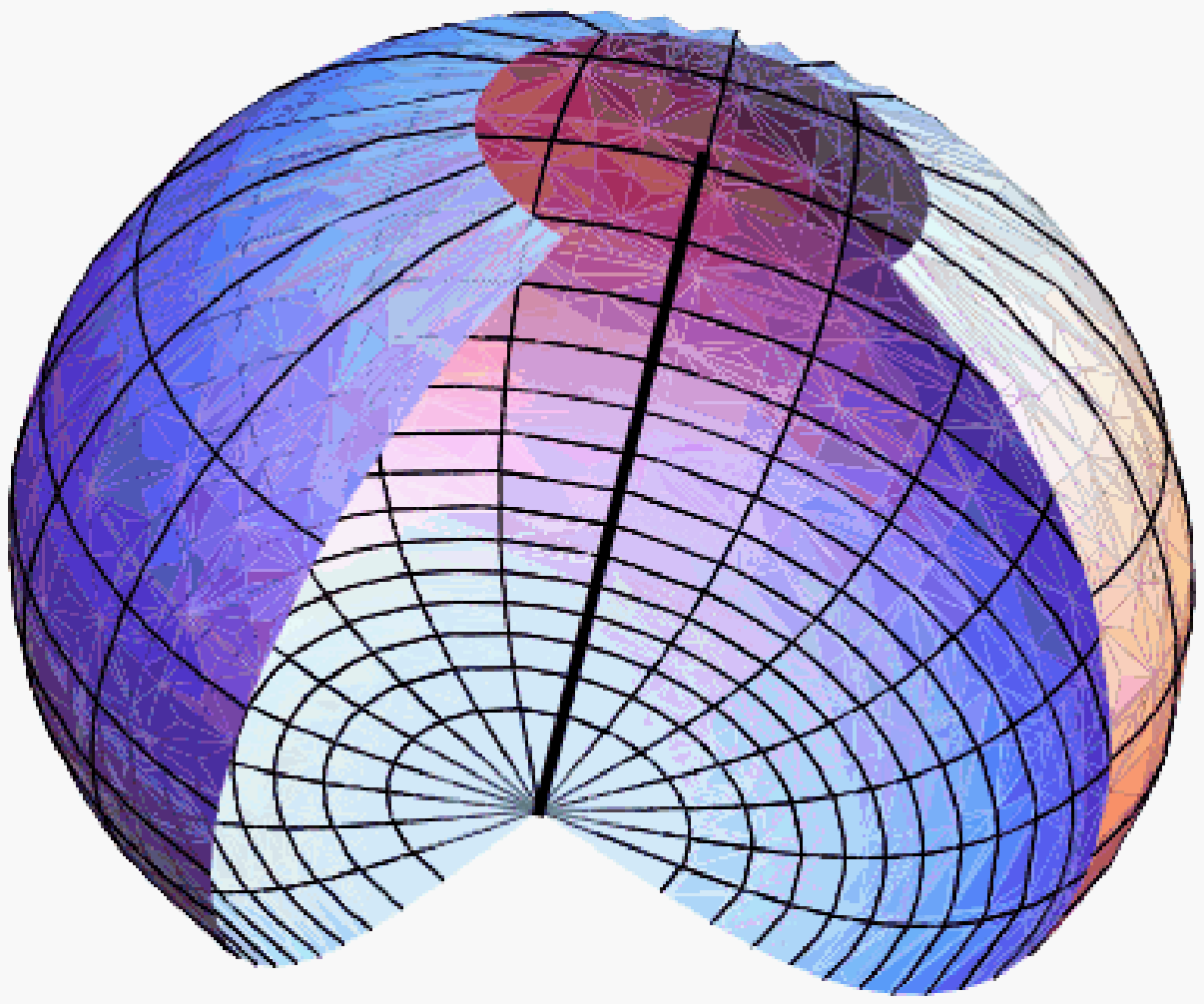}}
\centering\includegraphics[height=2in,width=2in]{{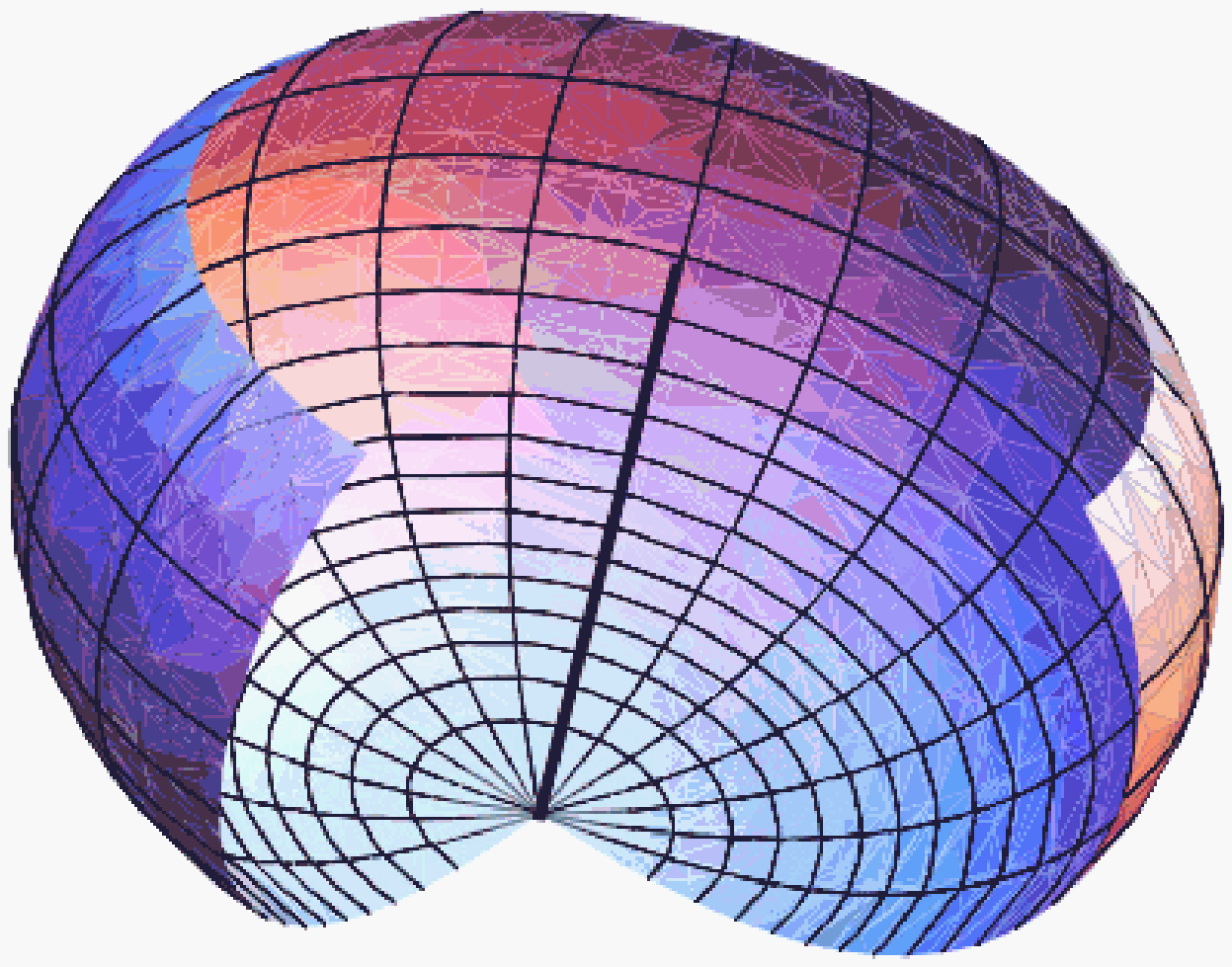}}
\centering\includegraphics[height=2in,width=2in]{{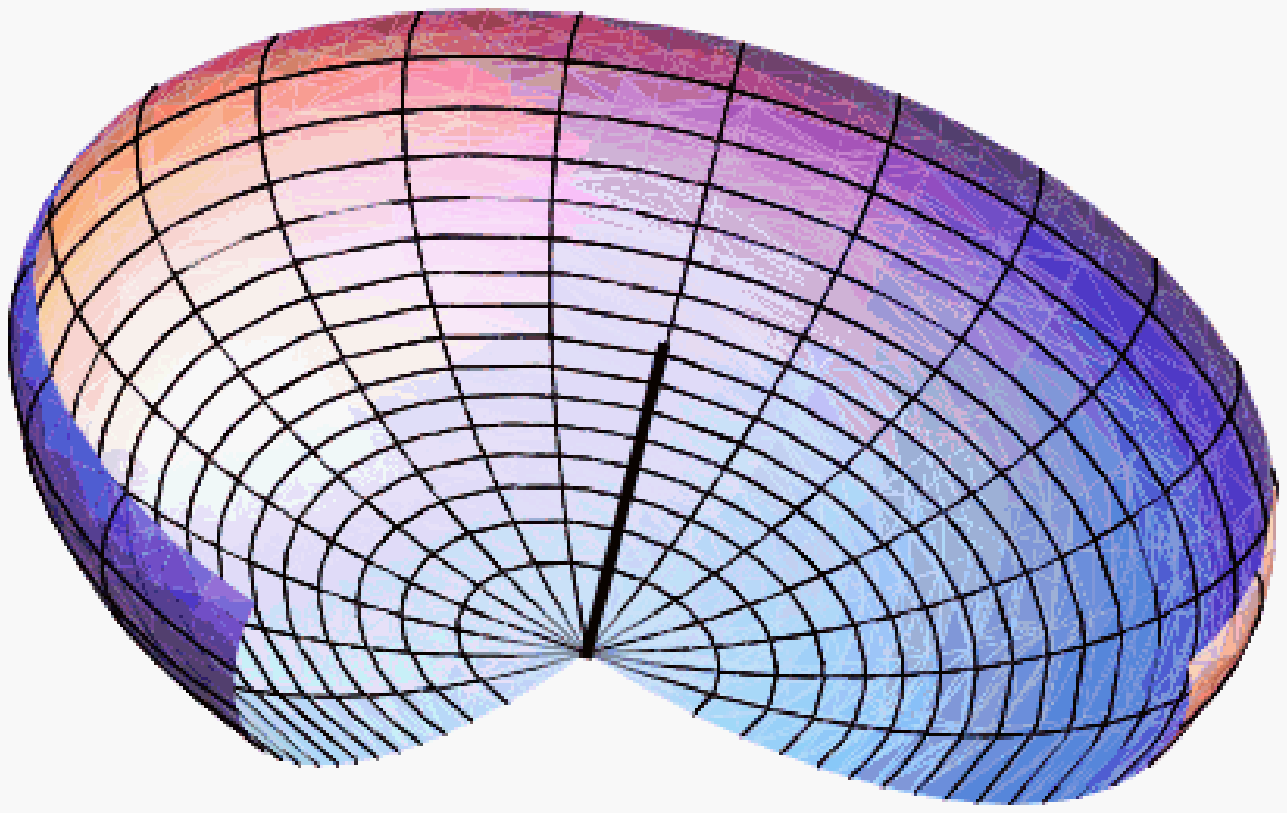}}
\begin{picture}(0,0)(0,0)
\end{picture}
\caption{Embedding the double Kerr horizon for fixed mass and angular momentum $M=1=J$ and various values of the physical distance $d$ ($d=25,10,3$) in $H^3$, with $k=1$. Unlike the Euclidean embedding, the south pole is always covered by the hyperbolic embedding. As in the Euclidean embedding, there is a growing patch around the north pole that is not covered by the hyperbolic embedding, manifesting the deformation induced by the strut.  The patch does not appear smaller than in the Euclidean embedding.}
\label{dkh}
\end{figure}

In Fig. \ref{dke} and \ref{dkh} we present 3D plots of the embeddings in $\mathbb{E}^3$ and $H^3$ of the ``lower" black hole in the counter-rotating double Kerr system.

\section{Conclusions and final remarks}
In this paper we have discussed global and unique isometric embeddings of 2-surfaces in hyperbolic 3-space. Such embeddings are possible as long as the Gaussian curvature of the 2-surface is bounded below. As an example we considered the embedding for the Kerr-Newman black hole event horizon. According to a fundamental theorem of Riemannian geometry \cite{Nash}, every smooth $n$-dimensional Riemannian manifold can be globally isometrically embedded in an N-dimensional Euclidean space, where $N=(n+2)(n+3)/2$.\footnote{A \textit{local} embedding, is possible in a lower dimensional Euclidean space, $\mathbb{E}^N$, with $N=n(n+1)/2$, according to a theorem first proved by Janet for 2-manifolds \cite{Janet} and subsequently generalised by Cartan for $n$-dimensional manifolds \cite{Cartan}. Note that the dimension of the embedding space is the number of components of the metric tensor of the manifold.} For $n=2$ this yields $N=10$. An embedding in a lower dimensional space may, nevertheless, be possible. Indeed it was shown by Frolov that, for the Kerr-Newman case, a global embedding in $\mathbb{E}^4$ is possible. Herein, we have shown that a global and unique embedding in a 3-dimensional \textit{hyperbolic} space is possible for all values of the angular momentum, in contrast with the embedding in Euclidean 3-space, first described in \cite{Smarr}, which is only global for $J/M^2\le \sqrt{3}/2$. Moreover we have shown that up to  $J/M^2\simeq 0.873> \sqrt{3}/2$, the hyperbolic embedding can be fitted in a fundamental domain of the Picard group, which is used to construct interesting quotients of hyperbolic space.

Recently, various physical properties of the double-Kerr system were studied \cite{Herdeiro:2008kq,Costa:2009wj}. One novel feature unveiled in these studies is that the angular velocity of the two black holes decreases as they are approached, keeping their mass and angular momentum fixed, for both the counter-rotating \cite{Herdeiro:2008kq} and co-rotating \cite{Costa:2009wj} cases. This effect was interpreted as a consequence of the mutual rotational dragging of the black holes. The angular velocity decrease is visible in the horizon geometry, and therefore in its embedding in a higher dimensional space. Although the effect of the angular velocity decrease is clearer in the Euclidean embedding, it is also visible in the hyperbolic embedding exhibited herein. Due to the existence of a strut which provides the force balance between the black holes, there is a curvature singularity at one of the poles of the horizon and therefore the embedding (both Euclidean and hyperbolic) is not global.

Other possible applications of the hyperbolic embedding are to Kerr-(Anti) de Sitter black holes and to  
 a Schwarzschild black hole immersed in a magnetic field. In the latter case it has been shown \cite{WildKerns} that a patch of negative Gaussian curvature develops, near the equator, if the magnetic field parameter exceeds a certain limit. Curiously, for a certain range of magnetic field values, the Gaussian curvature of a Kerr black hole immersed in a magnetic field becomes positive for all values of the angular momentum, and an embedding in $\mathbb{E}^3$ is possible \cite{KulkarniDadhich}.

\section*{Acknowledgements}
C.H. would like to thank the hospitality of D.A.M.T.P., University of
Cambridge, where part of this work was done. C.H. is supported by a ``Ci\^encia 2007" research contract. C.R. is funded by FCT through grant SFRH/BD/18502/2004

\end{document}